\documentclass[final,3p,times,11pt]{elsarticle}

\usepackage{amsmath,amssymb,amsfonts}
\usepackage{amsthm}
\usepackage[dvipsnames]{xcolor}
\usepackage{dsfont}

\usepackage[linesnumbered, ruled, vlined]{algorithm2e}
\usepackage{graphicx,subcaption,lipsum}
\usepackage{url}
\usepackage{doi}
\usepackage{hyperref}
\usepackage{placeins}
\usepackage{tabularx}
\newcolumntype{Y}{>{\centering\arraybackslash}X}
\usepackage{subcaption}
\usepackage[labelfont=bf]{caption}
\usepackage{float}
\usepackage{orcidlink}
\usepackage{enumitem}
\usepackage{natbib}
\usepackage{tikz}
\usepackage{pgfplots}
\usepackage{tabularx}
\newcolumntype{Y}{>{\centering\arraybackslash}X}
\usepackage{multirow}
\usepackage{url}

\usepackage{tikz}
\usetikzlibrary{tikzmark}

\renewcommand{\d}[1]{\textnormal{d}#1}
\usepackage{xr}
\usepackage[capitalize]{cleveref}
\creflabelformat{equation}{#2#1#3}

\journal{}

\begin{document}
	
	\begin{frontmatter}
		
		\title{A spatio-temporal hybrid approach for mathematical epidemiological metapopulation models: Combining stochastic ensemble simulations with moment-closure approximations}

		\author[label1]{Julia Bicker\corref{cor1}\hspace{0.5mm}\orcidlink{0000-0001-9382-4209}}
		\ead{julia.bicker@dlr.de}
		\cortext[cor1]{Corresponding author}
		\author[label2]{Kilian Volmer\hspace{0.5mm}\orcidlink{0009-0008-9861-0216}}
		\ead{kilian.volmer@uni-bonn.de}
		\author[label2]{Jan Hasenauer\hspace{0.5mm}\orcidlink{0000-0002-4935-3312}}
		\ead{jan.hasenauer@uni-bonn.de}
		\author[label1,label2]{Martin J. Kühn\hspace{0.5mm}\orcidlink{0000-0002-0906-6984}}
		\ead{martin.kuehn@dlr.de}
		
		\affiliation[label1]{organization={Institute of Software Technology, Department of High-Performance Computing, German Aerospace Center}, city={Cologne}, country={Germany}}
		
		\affiliation[label2]{organization={Bonn Center for Mathematical Life Sciences and Life \& Medical Sciences Institute, University of Bonn}, city={Bonn}, country={Germany}}
		
		\begin{abstract}
			Mathematical models support the evaluation of interventions to mitigate disease spread and inform public health decision-making. However, capturing stochastic epidemic dynamics across connected populations can be computationally demanding. While moment equations offer an efficient alternative to stochastic ensemble simulations, their accuracy depends on the dynamical regime. Hybrid models are a promising tool for combining the efficiency of moment-closure approximations with the detail of stochastic ensemble simulations, thereby balancing the trade-off between accuracy and computational complexity. In this work, we present a spatio-temporal hybrid framework in which each region of a metapopulation model switches independently between representations, guided by ensemble-level statistics. We systematically assess the accuracy of the moment-closure approximation and evaluate how switching criteria, initial conditions, and model parameters affect hybrid accuracy and runtime. An application to seasonal influenza-like illness dynamics in Germany illustrates the framework's use for regional intervention scenarios, with runtime reductions of up to $95\%$ relative to the stochastic benchmark.
			
		\end{abstract}
		
		
	\end{frontmatter}
	
	
	\section{Introduction}\label{sec:introduction}
	
	Mathematical models have become the principal tool for predicting and assessing the dynamics of epidemic outbreaks~\cite{ecdc_npis_2024}. Modeling approaches can be broadly categorized based on their resolution and dynamics. Population-based mean-field models describe the numbers or proportions of individuals in distinct epidemiological states~\cite{khailaie_development_2021,bjornstad_seirs_2020,giordano_modelling_2020}. Metapopulation models extend this representation by coupling subpopulations to account for spatial and demographic heterogeneity~\cite{ArinoVdD2003a,COSNER2009550,wang2018,Citron2021,liu_modelling_2022,bicker2026memilio,ZUNKER2027_stage_aligned}, whereas agent-based models explicitly represent individuals and their interactions~\cite{KERKMANN_ABM_2025,kerr_covasim_2021,shattock_impact_2022,chao_flute_2010}. Dynamics can be formulated deterministically or stochastically. Deterministic population-based and metapopulation models, commonly formulated as ordinary differential equations (ODEs), are widely used because of their computational efficiency and established methods of analysis. While their simplest formulations assume homogeneous mixing, more complex formulations can account for heterogeneous contact patterns between population groups~\cite{kuhn_assessment_2021,rozhnova_model-based_2021}. However, deterministic models neglect random fluctuations in infection and recovery events, which can strongly influence epidemic dynamics, especially when only a small number of individuals is infected. Stochastic formulations capture these fluctuations at both the individual and the population level, but often entail higher computational costs.
	
	Continuous-time Markov chains (CTMCs) provide a flexible framework for stochastic epidemic modeling~\cite{krishnarajah_novel_2005,winkelmann_mathematical_2021,achterberg_sis_2022}. Agent-based models with Markovian transition rules can be formulated as CTMCs and, when transition rates depend only on population counts, recast as population-based or metapopulation CTMCs. Non-exponentially distributed dwell times, such as infectious periods, can also be represented or approximated by introducing auxiliary stages using the linear chain trick and its generalizations~\cite{hurtado_generalizations_2019}. This retains the Markov property in the enlarged state space while allowing flexible dwell-time distributions within epidemiological compartments~\cite{ploetzke_lct_2025}. Individual trajectories can be generated using exact stochastic simulation algorithms, such as the Gillespie algorithm~\cite{gillespie_exact_1977}. However, the computational cost grows at least linearly with the number of simulated events. Consequently, (repeated) simulations of large populations can become computationally demanding.
	
	While individual CTMC trajectories are stochastic, the corresponding probability distribution is governed by the master equation (ME), an ODE system describing the probability of being in a specific state at a given time. As the number of possible configurations grows rapidly with population size and model complexity, solving the ME quickly becomes computationally intractable. Instead, the method of moments (MoM) can be used to derive ODEs for statistical properties of this distribution, such as means, variances, and covariances~\cite{mcquarrie1967stochastic,engblomComputingmoments2006,lee_moment_2009}. For nonlinear transition propensities, lower-order moment equations generally depend on higher-order moments, requiring moment-closure approximations to obtain a tractable system. Various closure schemes have been proposed, including truncation~\cite{gillespie_moment-closure_2009}, distributional closures~\cite{krishnarajah_novel_2005,singh_lognormal_2006,bronstein_variational_2018}, derivative matching~\cite{singh_derivative_2007,singh_approximate_2011}, and conditional closures~\cite{hasenauer_method_2014}. The resulting moment equations provide a computationally efficient approximation of the statistical properties of stochastic epidemic dynamics. However, their accuracy depends on the chosen closure scheme and the dynamical regime. Moreover, a finite set of low-order moments does not generally determine the full probability distribution or the statistics of individual trajectories, limiting the direct assessment of outcomes such as epidemic extinction.
	
	The prior considerations highlight that each modeling approach has distinct advantages and inherent limitations, thus choosing an appropriate model depends on the specific epidemiological context, population structure, and desired level of detail, with implications for both accuracy and computational cost. Hybrid modeling approaches aim to balance these requirements by combining different model formulations across space and time. Spatial hybridization couples formulations assigned to different spatial regions. For example, Yoneyama et al.~\cite{yoneyama_hybrid_2012} developed a hybrid SARS model that couples local ODE-based models with a global infection layer to account for inter-region traffic. Related approaches were suggested by Kasereka et al.~\cite{kasereka_hybrid_2014}, who considered different reproduction numbers for a network of cities, and Bradhurst et al~\cite{bradhurst_hybrid_2015}, who simulated disease spread on a graph of livestock herds. While these models combine aspects of ODE- and network- or agent-based modeling, they still use population-based modeling assumptions and are thus similar to (graph-based) metapopulation approaches~\cite{ZUNKER2027_stage_aligned}. Other spatial hybrid approaches combine agent-based and ODE-based models~\cite{bostanci2025}, PDE- and ODE-based models~\cite{kehrer_hybrid_2025}, agent-based and PDE-based models~\cite{kehrer_hybrid_2027}, or empirically derived contact networks for mass gathering events with stochastic metapopulation models~\cite{kwak2026hybridepidemicsimulationframework}.
	
	Temporal hybridization instead switches between formulations as the epidemic evolves. Bobashev et al.~\cite{bobashev_hybrid_2007} proposed switching between a stochastic agent-based and a deterministic ODE-based model according to a case number threshold.  Based on the idea that stochastic agent-based models are more appropriate to simulate disease dynamics in low incidence regimes than population-based models, this approach retains an individual-level stochastic description when case numbers are low and uses a computationally efficient deterministic description when case numbers are sufficiently large. Hunter et al.~\cite{hunter_hybrid_2020} investigated a related approach for disease spread at town and county levels.
	
	Recently, we presented a general framework for spatial and temporal hybridization of agent- and\linebreak metapopulation-based models~\cite{bicker_hybrid_2025}, using model formulations originally introduced in~\cite{winkelmann_mathematical_2021}. In the spatial hybridization, an agent-based model describes disease spread in a focus region, while ODEs describe dynamics in connected regions. In the temporal hybridization, the system switches between stochastic agent-based and deterministic population-based descriptions during the simulation. The resulting hybrid models reproduced key features of the agent-based dynamics, including epidemic extinction, while reducing runtime and energy consumption by up to 98~\%. These results illustrate the potential of hybrid modeling. However, determining how, where and when to switch between formulations remains a central methodological challenge. In particular, switching criteria must balance computational savings against the accuracy required to capture stochastic epidemic dynamics.
	
	In this study, we combine the distinct spatial and temporal hybridization approaches introduced in~\cite{bicker_hybrid_2025} in a novel spatio-temporal hybrid model and apply this framework to a continuous-time Markov model and the corresponding system of mean and moment equations. We first analyze the runtime scaling of both the moment-closure approximation and the stochastic model with respect to the population size and the number of modeled regions, quantifying the computational advantage of the moment equations. Second, we systematically investigate the accuracy of the moment-closure approximation under varying initial conditions. This analysis enables the derivation of criteria that determine when the moment-closure approximation is sufficiently accurate and when the stochastic representation is required. We then apply four distinct conditions to spatio-temporal hybrid SIR- and SIRS-type metapopulation models and evaluate the switching behavior, the accuracy relative to a fully stochastic reference, and the strong scaling behavior. Finally, we use the spatio-temporal hybrid model to simulate the spread of influenza-like illness (ILI) in Germany from August 2016 to July 2021, assessing the impact of various non-pharmaceutical interventions (NPIs) as decreed for mitigating COVID-19 from March 2020 onward. The results demonstrate the hybrid model's ability to accurately capture local outbreaks as well as long-term epidemic dynamics, highlighting its potential for epidemic forecasting and scenario analysis.
	
	\section{Materials and Methods}
	
	\subsection{Stochastic metapopulation model}\label{sec:Stochastic_model}
	In this manuscript, we consider stochastic metapopulation models of infectious disease dynamics formulated as continuous-time Markov chains. Given $n_Z$ epidemiological states and $n_R$ spatial regions, resulting in $n_C=n_Z\cdot n_R$ subpopulations (or compartments), the system state is given by the stacked vector
	\begin{align*}
		X(t)=\left(X^{(1)}(t)^\top,\ldots,X^{(n_R)}(t)^\top\right)^\top\in\mathbb{N}_0^{n_C},
	\end{align*} 
	where 
	\begin{align*}
		X^{(k)}=\left(X_1^{(k)},\ldots,X_{n_Z}^{(k)}\right)^\top
	\end{align*}
	records the numbers of individuals in region $k\in\{1,\ldots,n_R\}$ and $X_z^{(k)}(t)$ denotes the number of individuals in region $k$ having epidemiological state $z\in\{1,\ldots,n_Z\}$. For readability, we also use an equivalent flat indexing, $X(t)=\left(X_1(t),\ldots,X_{n_C}(t)\right)^\top$, and an index mapping $\iota(z,k)=(k-1)n_Z+z$ with
	\begin{align*}
		X_{\iota(z,k)}=X_z^{(k)}.
	\end{align*} 
	The dynamics are specified by $n_T$ transition types, describing changes in epidemiological state or movement between regions. Each transition type $r\in\{1,\ldots,n_T\}$ has a (time-dependent) propensity $a_r:\mathbb{N}_0^{n_C}\times\mathbb{R}_{\geq0}\rightarrow[0,\infty)$ and a change vector $\nu_r\in\mathbb{Z}^{n_C}$, whose $i$-th entry is the net change in subpopulation $i\in\{1,\ldots,n_C\}$ for transition event $r$. When an event of type $r$ occurs, the system state changes from $x$ to $x+\nu_r$. The propensity $a_r(x,t)$ determines the rate of this event in state $x$ at time $t$. While individual trajectories are stochastic, the probability distribution $P(x,t):=\mathbb{P}(X(t)=x)$ evolves according to the master equation,
	\begin{align}
		\frac{\d}{\d t}P(x,t)
		= \sum_{r=1}^{n_T}
		\left[
		a_r(x-\nu_r,t)P(x-\nu_r,t)
		- a_r(x,t)P(x,t)
		\right],
		\label{eq:ME}
	\end{align}
	which is a system of coupled ODEs. Terms associated with states outside the admissible state space are taken to be zero.
	
	We sample individual trajectories using exact stochastic simulation algorithms~\cite{gillespie_exact_1977,andersonmodifiednext2007}. These algorithms generate successive transition events, with a computational cost that grows with the number of simulated events and the cost of selecting each event and updating the affected propensities. Consequently, simulations can become computationally demanding for large populations, fast transitions, or ensembles requiring many realizations. This motivates the use of moment equations which describe the evolution of moments of the CTMC’s state distribution, together with moment-closure approximations, which yield a tractable system of equations.
	
	\begin{figure}[!h]
		\centering
		\includegraphics[width=\linewidth]{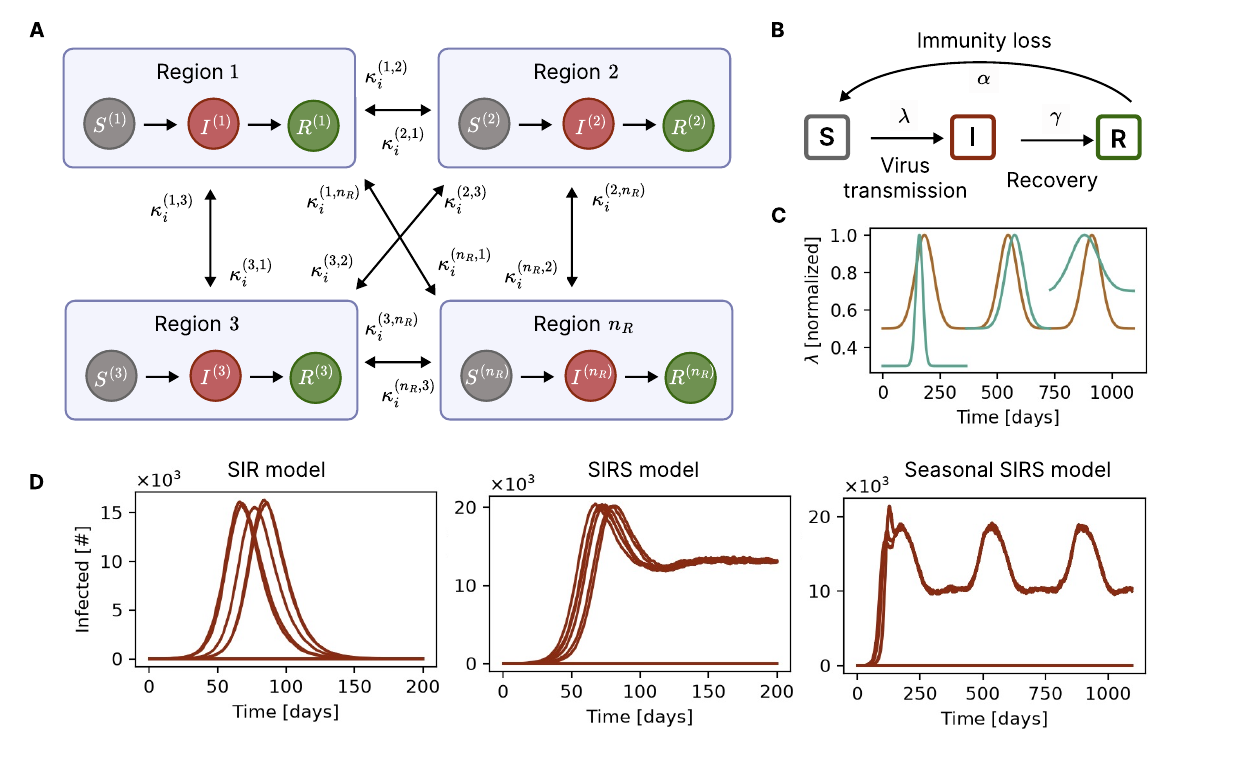}
		\caption{\textbf{Overview of the stochastic model.} \textbf{A:} Susceptible-Infected-Recovered (SIR) metapopulation model with spatial transition rates $\kappa_i^{(k,l)}$, $i\in\{S,I,R\}$. \textbf{B:} Flow chart of the SIRS type epidemiological model. \textbf{C:} Seasonal variation of the transmission rate $\lambda$ given a normalized peak rate of $\lambda_0=1$. One setting with identical $\rho_s=0.6, \sigma_s=40$ in every season and $t_s=365\cdot s+182$, $s\in\{0,1,2\}$ (brown), and one setting with varying parameters per season ($(\rho_0,\rho_1,\rho_2)=(0.3,0.5,0.7)$, $(t_0,t_1,t_2)=(160,365 + 210,2*365 + 150),$ $(\sigma_0,\sigma_1,\sigma_2)=(15,40,60)$) (teal). \textbf{D:} Exemplary trajectories for the SIR (left), the SIRS (middle) and the seasonal SIRS (right) model. Shown are $10$ runs with identical parameters and initial conditions for each model.}
		\label{fig:overview_stochastic}
	\end{figure}
	
	\subsection{Method of moments (MoM)}
	The MoM provides a system of ODEs describing the time evolution of mean, (co)variance, and higher-order (central) moments of the solution of the ME (\cref{eq:ME}). For CTMCs whose propensities are polynomial of degree at most two in the state variables, with possibly time-dependent coefficients, the derivation from~\cite{lee_moment_2009} can be employed. Consequently, the mean $\mu=(\mu_{1},\ldots,\mu_{n_C})^\top$ is given by
	\begin{align*}
		\frac{d}{dt}\mu_{c}=\sum_{r=1}^{n_T}\nu_{r,c}\biggl(a_r(\mu,t)+\frac{1}{2}\sum_{i,j=1}^{n_C}\frac{\partial^2a_r(\mu,t)}{\partial X_i\partial X_j}\sigma_{i,j}\biggr)
	\end{align*}
	where $\nu_{r,c}$ denotes the $c$-th entry of $\nu_{r}$. The mean number of individuals in subpopulation $c=1,\ldots,n_C$ depends on the (co)variances between subpopulations $X_i$ and $X_j$, $\sigma_{i,j}=\mathbb{E}\left[(X_{i}-\mu_{i})(X_{j}-\mu_{j})\right]$, if there exist $i$ and $j$ such that $\frac{\partial^2a_r(\mu,t)}{\partial X_i\partial X_j} \neq 0$. Hence, the means can depend on a set of central moments of order two. The governing equations for variances and covariances are given by
	\begin{align*}
		\frac{d}{dt}\sigma_{i,j}&=\sum_{r=1}^{n_T}\biggl[\sum_{k=1}^{n_C}\frac{\partial a_r(\mu,t)}{\partial X_k}\bigg(\nu_{r,i}\sigma_{j,k}+\nu_{r,j}\sigma_{i,k}\biggr)+\nu_{r,i}\nu_{r,j}\biggl(a_r(\mu,t)+\frac{1}{2}\sum_{k,l=1}^{n_C}\frac{\partial^2a_r(\mu,t)}{\partial X_k\partial X_l}\sigma_{k,l}\biggr)\\
		&+\frac{1}{2}\sum_{k,l=1}^{n_C}\frac{\partial^2a_r(\mu,t)}{\partial X_k\partial X_l}\biggl(\nu_{r,i}\mathbb{E}[(X_j-\mu_j)(X_k-\mu_k)(X_l-\mu_l)]+\nu_{r,j}\mathbb{E}[(X_i-\mu_i)(X_k-\mu_k)(X_l-\mu_l)]\biggr)\biggr].
	\end{align*}
	Similar to the equations for mean and (co)variances, a general system of ODEs governing all central moments can be derived according to~\cite{lee_moment_2009}.
	
	For systems involving quadratic or higher-order propensities, the moment equations of a given order may depend on moments of higher order, thus the resulting ODE system is not necessarily closed. Therefore, moment closure schemes that approximate higher-order moments in terms of lower-order ones are required to render the system solvable~\cite{hespanha_moment_2008,schnoerr_comparison_2015}. While this introduces approximation error, the resulting moment-closure approximations can be computed efficiently using standard ODE solvers, providing a computationally efficient alternative to estimating moments from stochastic simulation ensembles. In this study, the Boost implementation~\cite{Ahnert_boost} of the adaptive Runge-Kutta-Cash-Karp54 method is used to solve ODE systems.
	
	Although population size can influence the dynamics and thus the computational cost of solving the moment equations, the number of state variables remains unchanged. Consequently, the computational cost typically depends much weaker on population size than that of stochastic simulation algorithms, but scales with the number of ODEs governing the system which depends on the number of compartments.
	
	\subsection{Hybrid modeling approach}\label{sec:hybrid_modeling_approach}
	\begin{figure}[!ht]
		\centering
		\includegraphics[width=\linewidth]{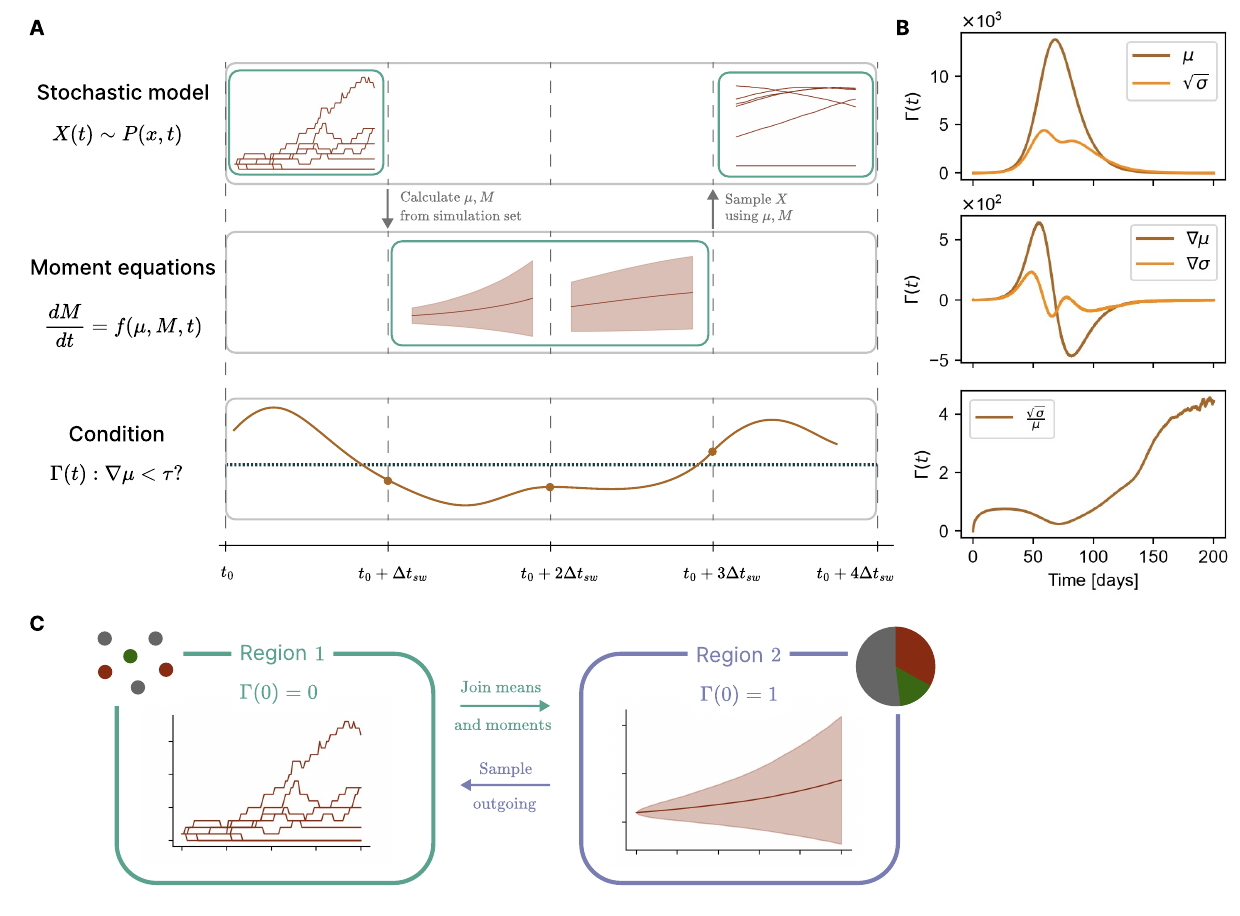}
		\caption{\textbf{Overview of the hybrid modeling approaches.} \textbf{A:} Temporal hybridization. The condition indicates which model is used and is evaluated at discrete time points in intervals $\Delta t_{sw}$. In the visualization, the moment-closure approximation is used if the mean gradient is below a threshold $\tau$ and Gillespie simulations are used if the mean gradient is above $\tau$. \textbf{B:} Possible quantities used for a switching condition. Shown are the mean and standard deviation (top), their gradients (middle) and the relation of current mean and standard deviation (bottom) over time. \textbf{C:} Spatial hybridization. Region~$1$ uses stochastic simulations and region~$2$ uses the moment-closure approximation for simulation. Upon spatial transitions, the corresponding system states are adapted.}
		\label{fig:hybrid_model_overview}
	\end{figure}
	To combine the computational efficiency of the MoM with the accuracy of stochastic Gillespie simulations in regimes where the moment-closure approximation fails, we introduce a stochastic-deterministic hybrid modeling approach. Based on~\cite{bicker_hybrid_2025}, the hybrid model employs two types of hybridization: A temporal hybridization and a spatial hybridization.
	
	For both approaches, we define two condition functions:
	\begin{itemize}
		\item $\Gamma^{(S\rightarrow D)}: \mathbb{N}_0^{n_{C}\times n_{sims}} \rightarrow\{0,1\}$, which maps the current system state of $n_{sims}$ independent stochastic simulations to a binary value and returns $1$ if the joint system state satisfies a predefined criterion for switching to the moment-closure approximation, and 0 otherwise.
		\item $\Gamma^{(D\rightarrow S)}: \mathbb{R}^{n_C+n_m} \rightarrow\{0,1\}$, which similarly maps the current MoM state, i.e., the solution of $n_C$ mean and $n_m=\binom{m+n_C}{m} - n_C - 1$ central moment equations for moments up to order $m$, to a binary value indicating if a switch to the stochastic model is required.
	\end{itemize}
	Depending on the output of the condition functions, the system state is evolved either using stochastic simulations or the moment-closure approximation. This adaptive strategy ensures high fidelity in time frames or regions where an exact stochastic representation is necessary while maintaining efficiency in regimes where the moment-closure approximation is sufficient.
	
	For a stochastic simulation $s\in\{1,\ldots,n_{sims}\}$, let $X_{z,s}^{(k)}$ be the number of individuals in region $k$ having epidemiological state $z$. The means of a stochastic ensemble of size $n_{sims}$ are given by
	\begin{align}\label{eq:stochastic_mean}
		\mu^S_{\iota(z,k)}=\frac{1}{n_{sims}}\sum_{s=1}^{n_{sims}}X_{z,s}^{(k)},
	\end{align}
	and the (co)variances by
	\begin{align}\label{eq:stochastic_var}
		\sigma^S_{\iota(z,k),\iota(y,l)}=\frac{1}{n_{sims}-1}\sum_{s=1}^{n_{sims}}\left(X_{z,s}^{(k)}-\mu^S_{\iota(z,k)}\right)\left(X_{y,s}^{(l)}-\mu^S_{\iota(y,l)}\right).
	\end{align}
	We denote the means and (co)variances obtained from the moment-closure approximation by $\mu^D_{\iota(z,k)}$ and $\sigma^D_{\iota(z,k),\iota(y,l)}$, respectively.
	
	In the temporal hybridization, the model switches dynamically between stochastic representation and moment-closure approximation during the simulation, hence $\Gamma^{(S\rightarrow D)}$ and $\Gamma^{(D\rightarrow S)}$ are evaluated in regular time intervals of size $\Delta t_{sw}$ (\cref{fig:hybrid_model_overview}A). This evaluation time step $\Delta t_{sw}$ restricts the maximum integrator step size $\Delta t_{max}$ used for solving the moment equations. If the switching condition is fulfilled, the whole system transitions from stochastic representation to moment-closure approximation, or vice versa. The condition is chosen in advance of the simulation and can depend on arbitrary statistics of the system state (\cref{fig:hybrid_model_overview}B). Converting from stochastic model to moment equations is straightforward: The current mean and central moments are computed directly from the ensemble of stochastic simulations (see~\cref{eq:stochastic_mean,eq:stochastic_var}) and used as initial conditions for the moment equations. In contrast, the conversion from moment equations to stochastic model requires reconstruction of the current state of the stochastic ensemble. This is achieved by sampling the current state for each of the $n_{sims}$ simulations from an appropriate probability distribution in line with the statistics given by the moment equations. In this study, we use the normal distribution specified in~\cref{eq:normal_dist} to sample values for the stochastic ensemble. Since our analysis focuses on ensemble-level statistics rather than individual simulation outcomes, we do not explicitly enforce simulation-level properties such as population conservation.
	
	For the spatial hybridization, different regions are modeled using distinct representations, based on a unique assignment at the beginning of the simulation governed by the condition function $\Gamma^{(S\rightarrow D)}$ (\cref{fig:hybrid_model_overview}C). If the moment-closure approximation is used in a region, means are set to initial compartment values, while moments of higher order are initialized with zero. To account for spatial transitions occurring between regions modeled with different model types, inter-model exchange of individuals is performed in regular intervals $\Delta t_{ex}$. This discrete inter-model exchange represents an approximation, as movement of individuals is in general not restricted to the model exchange time points. The approximation error depends on the choice of $\Delta t_{ex}$. Similar to $\Delta t_{sw}$ in the temporal hybridization, $\Delta t_{ex}$ is an upper limit for the maximum integrator step size. For the exchange from a region using moment equations to a stochastic region, the incoming group of individuals is sampled similar to the deterministic-stochastic conversion in the temporal hybridization. Precisely, for every simulation $s=1,\ldots,n_{sims}$, the number of (incoming) individuals in region~$k$ having epidemiological state $z$ is drawn according to
	\begin{align}\label{eq:normal_dist}
		\tilde{X}^{(k)}_{z,s} \sim \mathcal{N}(\mu^{D}_{\iota(z,k)}, {\sigma_{\iota(z,k),\iota(z,k)}^{D}}).
	\end{align}
	In the spatial hybridization, these statistics refer to the subpopulation entering region~$k$ from all regions modeled with moment equations during the last step of size $\Delta t_{ex}$. Since sampling from the normal distribution may yield non-integer or negative values, sampled values are adjusted via
	\begin{align} \label{eq:sampling_dist}
		X^{(k)}_{z,s} = \max\{0,\lfloor\tilde{X}^{(k)}_{z,s}+\xi_{\iota(z,k),s-1}\rceil\},
	\end{align}
	where $\xi_{\iota(z,k),s-1}$ is a cutoff and rounding error correction term intended to keep mean and variance of the sampled values close to the statistics obtained from the moment equations. $\lfloor\cdot\rceil$ denotes rounding to the nearest integer.
	
	When exchanging subpopulations from a stochastically modeled region to a region using moment equations, the ODE system state is updated by merging the mean and moments of the incoming stochastic subpopulation with the existing ones. For the mean, this update is done by employing linearity of expectation:
	\begin{align*}    
		\mu^{D}_{\iota(z,k)}\leftarrow\mu^{D}_{\iota(z,k)}+\mu^{S}_{\iota(z,k)}
	\end{align*}
	To maintain computational tractability, we do not consider covariance contributions from incoming stochastic populations, i.e., we neither update covariances upon exchange of subpopulations between models nor use them to update variances. Variances are updated using the general inequality
	\begin{align}\label{eq:var_inequality}
		\left(\sqrt{\mathrm{Var}[X]}-\sqrt{\mathrm{Var}[Y]}\right)^2 \leq \mathrm{Var}[X+Y] \leq \left(\sqrt{\mathrm{Var}[X]}+\sqrt{\mathrm{Var}[Y]}\right)^2.
	\end{align}
	Employing~\cref{eq:var_inequality}, the updated variance in a region~$k$ modeled with moment equations is sampled uniformly from
	\begin{align*}
		\tilde\sigma_{\iota(z,k),\iota(z,k)}^{D} &\sim U\biggl[\left(\sqrt{\sigma_{\iota(z,k),\iota(z,k)}^{D}}-\sqrt{\sigma_{\iota(z,k),\iota(z,k)}^{S}}\right)^2, \left(\sqrt{\sigma_{\iota(z,k),\iota(z,k)}^{D}}+\sqrt{\sigma_{\iota(z,k),\iota(z,k)}^{S}}\right)^2\biggr]\\
		\sigma_{\iota(z,k),\iota(z,k)}^{D}&\leftarrow\tilde\sigma_{\iota(z,k),\iota(z,k)}^{D}.
	\end{align*}
	The approximation above does not explicitly verify positive semi-definiteness of the complete covariance matrix. However, enforcing this property when reconstructing (co)variances is non-trivial and lies beyond the scope of this manuscript.
	
	The two introduced hybridization approaches are complementary rather than mutually exclusive and can therefore be combined in a spatio-temporal hybrid framework (Supplementary~\cref{alg:temporal_spatial_hybrid}). In that case, we choose $\Delta t_{switch}=\Delta t_{ex}$, ensuring that within-region model switching and the exchange of subpopulations between populations of different representations occur at identical time intervals.
	
	\subsection{Models for simulation study}
	In this study, we specialize the general CTMC formulation underlying the ME (\cref{eq:ME}) and consider three models of increasing complexity: (i)~an SIR model describing infection and recovery, (ii)~an SIRS model additionally accounting for waning immunity, and (iii)~a seasonal SIRS model with time-dependent transmission rates. All three models describe a population distributed across $n_R$ regions. Within each region, individuals are classified as \textit{susceptible} ($S$), \textit{infected} ($I$), or \textit{recovered} ($R$), resulting in $n_C=3n_R$ compartments (\cref{fig:overview_stochastic}A). Therefore, the system state is $X(t)=\left(X^{(1)}(t)^\top,\ldots,X^{(n_R)}(t)^\top\right)^\top\in\mathbb{N}_0^{3 n_R}$ with
	\begin{align}
		X^{(k)}(t)=\left(X^{(k)}_S(t),\,X^{(k)}_I(t),\,X^{(k)}_R(t)\right)^\top.
	\end{align}
	We neglect births and deaths, so the total population size $N$ remains constant over time, while regional population sizes can change through movement.
	
	All three models include movement between distinct regions $k$ and $l$, described by spatial transitions $z^{(k)}\rightarrow z^{(l)}$ with propensity $\kappa_z^{(k,l)}X^{(k)}_z(t)$. Here, $\kappa_z^{(k,l)}$ denotes the per-capita movement rate from region~$k$ to region~$l$ for individuals in epidemiological state $z\in\{S,I,R\}$. Movement preserves the individual's epidemiological state. The within-region transitions and their rates are specified below.
	
	\paragraph{SIR model}
	The SIR model includes virus transmission and recovery (\cref{fig:overview_stochastic}A). In region~$k$, virus transmission from infected to susceptible individuals is described by the transition $S^{(k)}\rightarrow I^{(k)}$ with propensity $\lambda^{(k)}X^{(k)}_S(t)X^{(k)}_I(t)$, and recovery from infection by $I^{(k)}\rightarrow R^{(k)}$ with propensity $\gamma X^{(k)}_I(t)$. The transmission parameter $\lambda^{(k)}$ can vary between regions to account for differences in contact frequency and patterns, whereas the recovery rate $\gamma$ is assumed to be identical across regions.
	
	\paragraph{SIRS model}
	The SIRS model extends the SIR model by allowing immunity to wane after recovery (\cref{fig:overview_stochastic}B), accounting for the possibility of reinfection in diseases such as influenza or COVID-19~\cite{bull_waning_2025,petrova_evolution_2018}. It includes the additional transition $R^{(k)}\rightarrow S^{(k)}$ with propensity $\alpha X^{(k)}_R(t)$, where the waning rate $\alpha$ is assumed to be identical across regions.
	
	\paragraph{Seasonal SIRS model}
	The seasonal SIRS model retains the transitions of the SIRS model and introduces time-dependent transmission rates to account for seasonal variations. While these variations in transmission dynamics of infectious diseases are not entirely understood, potential drivers include temperature, humidity, and changes in indoor contact patterns~\cite{moriyama_seasonality_2020}. Based on~\cite{Edlund2011}, we describe seasonality using a time-dependent transmission rate $\lambda:\mathbb{R}_{\geq 0}\rightarrow\mathbb{R}_{\geq 0}$ given by
	\begin{align}\label{eq:seasonality_lambda}
		\lambda(t) = \lambda_{0}\biggl(\rho_s+\left(1-\rho_s\right)\frac{g(t-t_s,w_s)}{g(0,w_s)}\zeta(t)\biggr),
	\end{align}
	where $\lambda_{0}\in\mathbb{R}_{>0}$ is the peak transmission rate, $\rho_s\in[0,1]$ the minimal transmission proportion in season $s\in\mathbb{N}_0$, $t_s\in\mathbb{R}_{\geq 0}$ the time point of peak transmission in season $s$ and $g(t, w_s)$ is a Gaussian distribution with standard deviation $w_s\in\mathbb{R}_{>0}$ and mean zero. We assume a season length of one year, thus $s=\left\lfloor\frac{t}{365}\right\rfloor$. $\zeta:\mathbb{R}_{\geq 0}\rightarrow\mathbb{R}_{\geq 0}$ is a mixing function defined as
	\begin{align*}
		\zeta(t)=\begin{cases}
			\frac{1}{60}(t-\tau_s^{-}+30), & \text{if } t<\tau_s^{-}+30\\
			1, & \text{if } \tau_s^{-}+30\leq t\leq\tau_s^{+}-30\\
			\frac{1}{60}(\tau_s^{+}+30-t), & \text{if } t>\tau_s^{+}-30
		\end{cases}
	\end{align*}
	$\tau_s^{-}=365s$ and $\tau_s^{+}=\tau_s^{-}+365$ denote the start and end time point of season $s$, respectively. Using identical seasonality settings across all seasons, i.e., $\rho_i=\rho_j$, $t_i=t_{i-1} + 365$, and $w_i=w_j$ for $i,j\in\mathbb{N}$ with given initial values $\rho_0$, $t_0$ and $w_0$, results in interchangeable temporal variation patterns of the transmission rate in each season. In contrast, varying the parameters between seasons generates season-specific transmission dynamics and leads to greater heterogeneity in the transmission rate across seasons (\cref{fig:overview_stochastic}C).
	
	Disease dynamics differ substantially among the SIR, the SIRS and the seasonal SIRS models. Although all three models capture virus extinction for an appropriate parameter choice in a single region, their behavior diverges when sustained transmission and outbreak events occur (\cref{fig:overview_stochastic}D). 
	Overall, we observed that the SIR model simulations produce, at maximum, one single epidemic wave after which infection dies out due to depletion of susceptible individuals. In contrast, given a sufficiently high transmission rate and fraction of initial susceptibles, in outbreak simulations the SIRS model initially exhibits an infection wave and subsequently converges to an equilibrium state, characterized by a persistent, non-negligible infection prevalence (\cref{fig:overview_stochastic}D). The seasonal SIRS model, however, may produce recurrent epidemic waves in a single region which are driven by the temporal variations in the transmission rate.
	
	Since the virus transmission transition $S^{(k)}\rightarrow I^{(k)}$ has a quadratic propensity, the moment equations for central moments of order two contain third-order central moments. To facilitate the numerical simulation, we therefore apply a truncation closure with $m=2$, i.e.,
	\begin{align*}
		\mathbb{E}\biggl[(X^{(1)}_S-\mu_{\iota(S,1)})^{i_S^{(1)}}\ldots(X^{(n_R)}_R-\mu_{\iota(R,n_R)})^{i_R^{(n_R)}}\biggr] = 0, \quad \forall (i_S^{(1)},\ldots,i_R^{(n_R)}): i_S^{(1)}+\ldots+i_R^{(n_R)}=3.
	\end{align*}
	Various alternatives to truncation closure exist, including closures based on distributional assumptions, however their accuracy depends on the model and parameter values (Supplementary~\cref{fig:supp_moment_closures}).
	The moment equations for the SIRS-type metapopulation model are provided in Supplementary~\cref{sec:supp_moment_eq}. The equations for the seasonal SIRS-type model follow by using the time-dependent transmission rate $\lambda(t)$ from~\cref{eq:seasonality_lambda} and those for the SIR-type model by setting the waning rate $\alpha=0$.
	
	\begin{figure}[!ht]
		\centering
		\includegraphics[width=\linewidth]{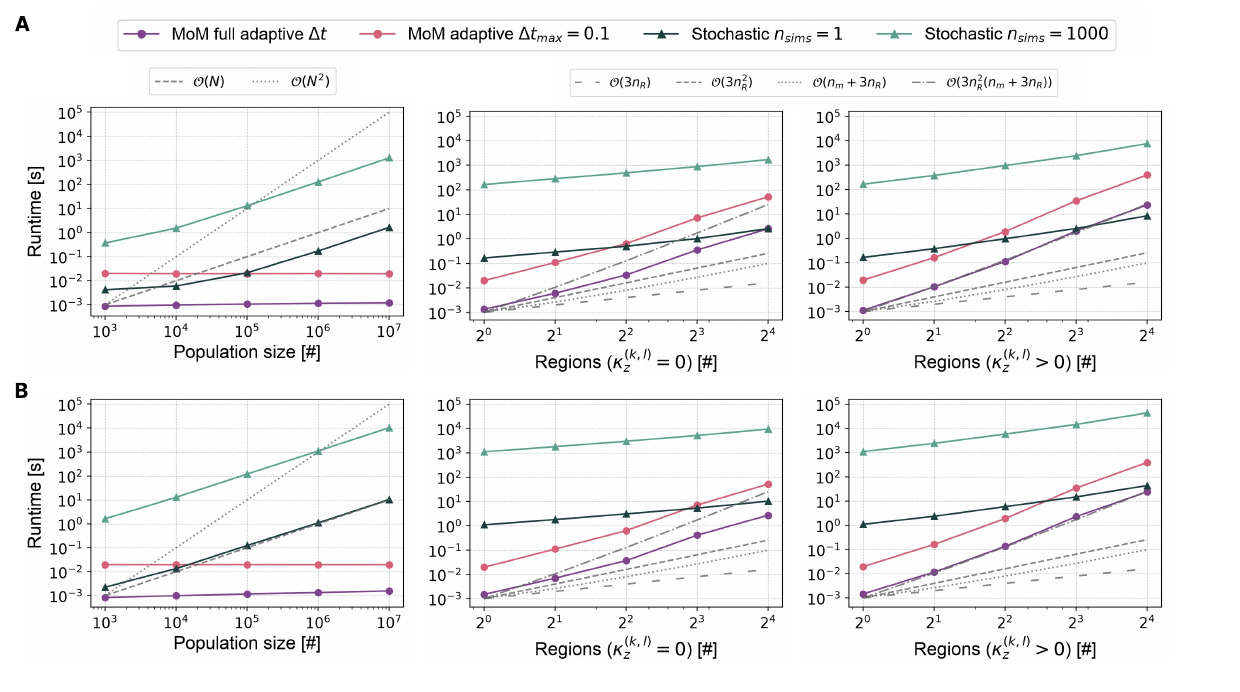}
		\caption{\textbf{Scaling behavior of the stochastic model and the moment-closure approximation for varying population sizes (left) and a varying number of regions without (middle) and with (right) spatial transitions between regions.} Shown are the results for sequential simulation on one core. For the stochastic model a set of $1$ and $1000$ simulations is shown. For the moment-closure approximation, we compare two settings: A fully adaptive integrator step size and an adaptive integrator step size constrained by a maximum step size of 0.1. Shown are the results for the \textbf{A:} SIR and the \textbf{B:} SIRS model.}
		\label{fig:scaling_MoM_SMM}
	\end{figure}
	
	\section{Results}
	\subsection{Runtime analysis of stochastic model and moment-closure approximation}
	
	To assess the challenge of using metapopulation models based on CTMCs for large-scale simulations, we compared the runtime scaling of stochastic ensembles and moment-closure approximations with respect to the population size and the number of regions. Therefore, we considered three settings for the SIR and SIRS-type models: (i)~A single region with a total population size ranging from $10^3$ to $10^7$ individuals and an initial reproduction number of $2.5$; (ii)~a growing number of $1$ to $16$ regions with a total population size of $10^6$, equally distributed across regions. The transmission rates were set such that the initial reproduction number in each region was $2.5$ and no spatial transitions were considered i.e. $\kappa_z^{(k,l)}=0$, $\forall$ $k,l=1,\ldots,n_R$, $z\in\{S,I,R\}$; (iii)~the same setting as in (ii) with inter-regional mobility incorporated by using spatial transition rates $\kappa_z^{(k,l)}=10^{-4}$ for $k\neq l$.
	
	The parameters and initial conditions for all settings of the runtime analysis can be found in Supplementary~\cref{tab:parameters_scaling_application}. We simulated the scenarios using the exact stochastic simulation algorithm (SSA)~\cite{andersonmodifiednext2007} as well as the MoM with truncation closure. For the SSA, we compared the total runtime of $1$ and $1{,}000$ simulations. For the MoM, we used an unbound maximum integrator step size $\Delta t_{max}$ as well as $\Delta t_{max}=0.1$.
	
	Our analysis for both models reveals that, on the one hand, the computational effort of the stochastic simulation increases with the number of modeled regions (\cref{fig:scaling_MoM_SMM}, middle and right, settings (ii) and (iii)), as every region introduces additional transitions whose propensities have to be computed and updated. On the other hand, since propensities depend on compartment sizes, computational cost of the stochastic simulation also scales with the total population size (\cref{fig:scaling_MoM_SMM}, left, setting (i)). 
	
	As the number of means and (central) moments depends on the number of modeled compartments, the computational cost of the MoM increases with the number of regions. Further, the evaluation of a given moment equation involves terms for spatial transitions (see Supplementary~\cref{eq:general_moment_eq}). Consequently, the MoM exhibits lower runtimes for spatially independent regions compared to a fully connected region network (\cref{fig:scaling_MoM_SMM}, middle and right, settings (ii) and (iii)). Additionally, the MoM runtime is influenced by the integrator step size. Allowing a fully adaptive step size reduces the MoM's runtime by approximately one order of magnitude compared to a constrained adaptive solver with a maximum step size of $\Delta t_{max} =0.1$ (\cref{fig:scaling_MoM_SMM}). This trade-off has to be considered when selecting $\Delta t_{ex}$, which is an upper bound for the maximum integrator step size in the hybrid model. Choosing the step size $\Delta t_{ex}$ too small reduces the computational advantage of the hybrid model compared to the fully stochastic model. Conversely, a too large value of $\Delta t_{ex}$ increases the approximation error due to a delayed incorporation of incoming subpopulations, leading to temporal inaccuracies.
	
	For a single region with a modeled population of $10$ million individuals, the MoM with a maximum step size of $\Delta t_{max} =0.1$ is $82$-times faster than a single run of the stochastic SIR-type model (setting (i)). When simulating an ensemble of $1{,}000$ independent stochastic simulations sequentially, a speedup of five orders of magnitude is achieved (\cref{fig:scaling_MoM_SMM}A, left). However, for a system with $16$ regions and a total population size of $1$ million -- resulting in an ODE system of $1{,}224$ coupled equations for the means and central moments of order two -- the MoM with $\Delta t_{max} =0.1$ and spatial transitions between regions (setting (iii)) is about $1.5$ orders of magnitude slower than a single stochastic run (\cref{fig:scaling_MoM_SMM}A, right). Nevertheless, for $1{,}000$ stochastic runs executed sequentially, the MoM reaches a speedup of $33$ and $19$ under the same step size constraint without (\cref{fig:scaling_MoM_SMM}A, middle, setting (ii)) and with (\cref{fig:scaling_MoM_SMM}A, right, setting (iii)) spatial transitions between regions, respectively. Due to the increased number of transmission and waning immunity transitions, the runtime advantage of the MoM over the stochastic model is even bigger for the SIRS-type model (\cref{fig:scaling_MoM_SMM}B). In that case, the MoM with bounded step size $\Delta t_{max} =0.1$ is about $99~\%$ faster than $1{,}000$ simulations of the stochastic model for $16$ regions and a total population of $1$ million (settings (ii) and (iii)), and even $5.5$ orders of magnitude for one region and a population of $10$ million (setting (i)).
	
	In summary, the analysis showed that for large population sizes and a small number of regions, the MoM is substantially faster than one individual CTMC simulation. However, when the number of regions is large, the MoM becomes slower than a single CTMC simulation due to the growing size of the ODE system as well as the increased number of spatial transition rates influencing each region. Nevertheless, since the MoM system needs to be solved only once, whereas reliable statistics from the CTMC usually require an ensemble of stochastic simulations, this additional cost is negligible as soon as various stochastic trajectories are needed.
	
	\subsection{Approximation performance of the MoM}
	\begin{figure}[!ht]
		\centering
		\includegraphics[width=0.91\linewidth]{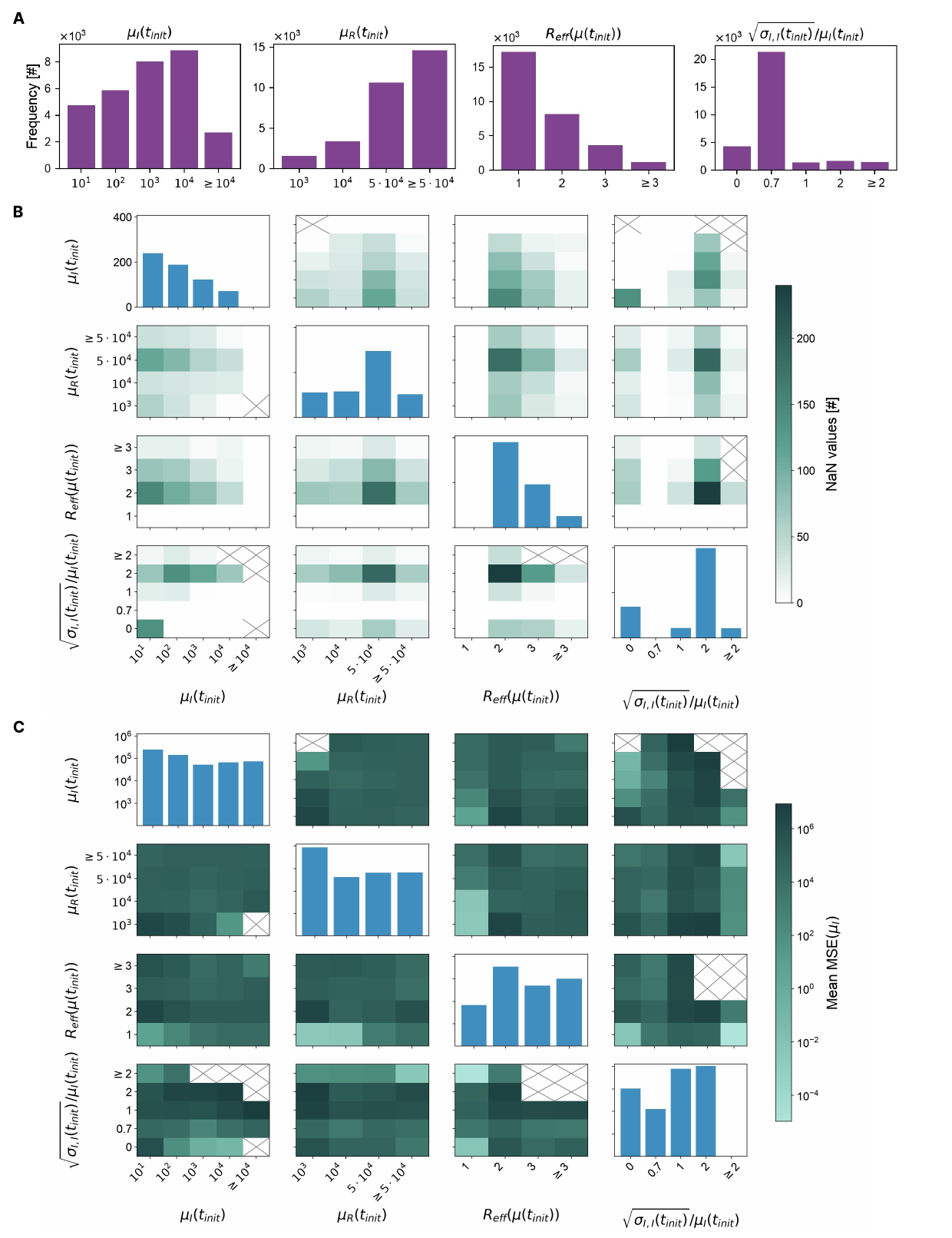}
		\caption{\textbf{Performance analysis of the MoM for the SIR model.} For the analysis, $50{,}000$ stochastic runs were compared to the solution of the moment equations using truncation closure for each parameter set. \textbf{A:} Distribution of initial conditions. \textbf{B:} Distribution of NaN value counts. \textbf{C:} Mean MSE of the simulated mean number of infected $\mu_I$. }
		\label{fig:parameter_study_SIR}
	\end{figure}
	To identify the configurations under which the MoM provides a good approximation to the underlying stochastic metapopulation model, we conducted a systematic parameter study for the SIR and the SIRS model, respectively (\cref{fig:parameter_study_SIR,fig:supp_parameter_study_SIRS}A). For each configuration, we compared the mean and variance of $50{,}000$ independent stochastic simulations to the corresponding output of the moment-closure approximation. We considered four quantities: the mean number of infected $\mu_I$ and recovered $\mu_R$, the standard deviation of the infected compartment $\sqrt{\sigma_{I,I}}$ and the reproduction number of the deterministic ODE-based compartmental model $R_{\textit{eff}}(\mu)=\frac{\lambda}{\gamma}\cdot\mu_S$. The moment-closure approximation fails, i.e. the solver returns NaN values, in scenarios with initial values of $\mu_I<10$ in combination with an initial standard deviation $\sqrt{\sigma_{I,I}}$ of zero or exceeding $0.7\mu_I$ (\cref{fig:parameter_study_SIR,fig:supp_parameter_study_SIRS}B). However, for configurations with an initial reproduction number of the mean $R_{\textit{eff}}(\mu(t_{init}))$ below one, the ODE system yields real-valued solutions for nearly all initial conditions. Additionally, the approximation error is minimal in this regime, as well as for initial conditions where $\frac{\sqrt{\sigma_{I,I}}}{\mu_{I}}<0.7$, indicating that the relative variability of the infected population is low (\cref{fig:supp_parameter_study_var,fig:parameter_study_SIR,fig:supp_parameter_study_SIRS}C). In contrast, the initial mean number of recovered does not exhibit a clear correlation with the accuracy of the moment-closure approximation.
	
	From this parameter study, we derived the condition functions $\Gamma^{(S\rightarrow D)}$ and $\Gamma^{(D\rightarrow S)}$ described in~\cref{sec:hybrid_modeling_approach} for the hybrid model. As the initial mean number of recovered does not show a clear influence on the accuracy of the moment-closure approximation, we based our conditions on the quantities $\mu_{I}$, $\frac{\sqrt{\sigma_{I,I}}}{\mu_{I}}$ and $R_{\textit{eff}}(\mu)$. For each of these, we defined separate conditions with corresponding thresholds or regimes where the moment-closure approximation performed well in our parameter study. We further introduced a fourth, combined condition that integrates all individual criteria. In addition to the quantity-dependent regimes, all condition functions use the moment-closure approximation, if the mean number of infectious individuals in a region falls below $1$, indicating a regime where infections are unlikely in the most stochastic simulations and the moment-closure approximation is considered accurate. With $\mathbf{X}\in\mathbb{N}_0^{n_C\times n_{sims}}$ the matrix containing the system state of $n_{sims}$ stochastic simulations, $\mu^D, \Sigma^D$ the vectors containing the means and central moments of order two of the moment-closure approximation and $R_{eff}^{(k)}(\mu)=\frac{\lambda^{(k)}}{\gamma}\mu_{\iota(S,k)}$, the four pairs of condition functions are defined as follows:
	\begin{itemize}
		\item An absolute threshold of $\tau_{\mu_{\iota(I,k)}}=10$ for the mean number of infected: 
		\begin{align*}
			\Gamma^{(S\rightarrow D)}(\mathbf{X})&=\begin{cases}
				1, & \text{if } \mu_{\iota(I,k)}^{S}>10 \lor \mu_{\iota(I,k)}^{S}<1\\
				0, & \text{otherwise}
			\end{cases}\\
			\Gamma^{(D\rightarrow S)}([\mu^{D}, \Sigma^D]^\top)&=\begin{cases}
				1, & \text{if } \mu_{\iota(I,k)}^{D}<10 \land \mu_{\iota(I,k)}^{D}\geq 1\\
				0, & \text{otherwise}
			\end{cases}
		\end{align*}
		\item A threshold of $1$ for the reproduction number of the mean $R_{eff}^{(k)}(\mu)$: 
		\begin{align*}
			\Gamma^{(S\rightarrow D)}(\mathbf{X})&=\begin{cases}
				1, & \text{if } R_{eff}^{(k)}(\mu^S)<1 \lor \mu_{\iota(I,k)}^{S}<1\\
				0, & \text{otherwise}
			\end{cases}\\
			\Gamma^{(D\rightarrow S)}([\mu^{D}, \Sigma^D]^\top)&=\begin{cases}
				1, & \text{if } R_{eff}^{(k)}(\mu^D)>1 \land \mu_{\iota(I,k)}^{D}\geq1\\
				0, & \text{otherwise}
			\end{cases}
		\end{align*}
		\item An interval of $(0,0.7)$ for the relation $\frac{\sqrt{\sigma_{\iota(I,k),\iota(I,k)}}}{\mu_{\iota(I,k)}}$ of standard deviation and mean of the infected compartment: 
		\begin{align*}
			\Gamma^{(S\rightarrow D)}(\mathbf{X})&=\begin{cases}
				1, & \text{if } \mu^{S}_{\iota(I,k)}\neq 0 \land\frac{\sqrt{\sigma^{S}_{\iota(I,k),\iota(I,k)}}}{\mu^{S}_{\iota(I,k)}} \in (0,0.7)\\
				0, & \text{otherwise}
			\end{cases}\\
			\Gamma^{(D\rightarrow S)}([\mu^{D}, \Sigma^D]^\top)&=\begin{cases}
				1, & \text{if } \mu^{D}_{\iota(I,k)}\neq 0\land \left(\frac{\sqrt{\sigma^{D}_{\iota(I,k),\iota(I,k)}}}{\mu^{D}_{\iota(I,k)}} \leq 0 \lor \frac{\sqrt{\sigma^{D}_{\iota(I,k),\iota(I,k)}}}{\mu^{D}_{\iota(I,k)}} > 0.7\right)\\
				0, & \text{otherwise}
			\end{cases}
		\end{align*}
		\item A combined condition integrating all previous ones:
		\begin{align*}
			\Gamma^{(S\rightarrow D)}(\mathbf{X})&=\begin{cases}
				1, & \text{if } \mu_{\iota(I,k)}^{S}<1 \lor R_{eff}^{(k)}\left(\mu^{S}\right)<1 \lor \left(\frac{\sqrt{\sigma^{S}_{\iota(I,k),\iota(I,k)}}}{\mu^{S}_{\iota(I,k)}} \in (0,0.7)\land \mu_{\iota(I,k)}^{S}>10\right)\\
				0, & \text{otherwise}
			\end{cases}\\
			\Gamma^{(D\rightarrow S)}([\mu^{D}, \Sigma^D]^\top)&=\begin{cases}
				1, & \text{if } \mu_{\iota(I,k)}^{D}\geq1 \land R_{eff}^{(k)}\left(\mu^{D}\right)>1 \land \left(\frac{\sqrt{\sigma^{D}_{\iota(I,k),\iota(I,k)}}}{\mu^{D}_{\iota(I,k)}} > 0.7 \lor \mu_{\iota(I,k)}^{D}<10\right)\\
				0, & \text{otherwise}
			\end{cases}
		\end{align*}
	\end{itemize}
	
	In summary, the parameter study identifies regimes in which the moment-closure approximation accurately reproduces stochastic ensemble statistics, particularly when the reproduction number is below one or relative infection variability is low. These findings motivate four condition functions based on mean infection levels, relative variability, and the reproduction number, which we evaluate in the hybrid model in the following section.
	
	\subsection{Approximation and runtime performance of the hybrid model} 
	
	\begin{figure}[!ht]
		\centering
		\includegraphics[width=\linewidth]{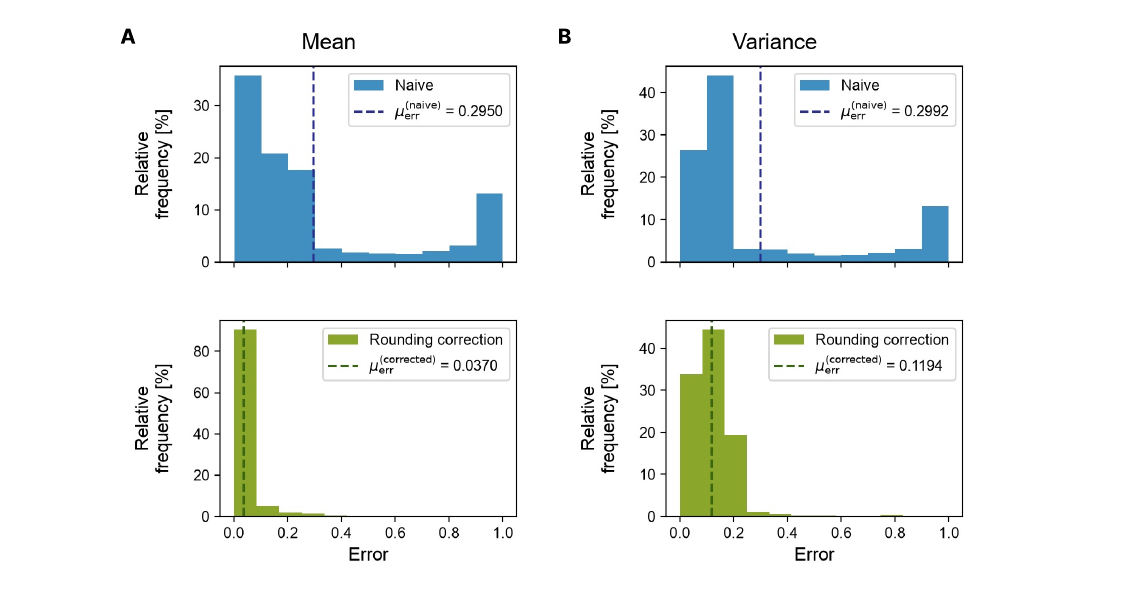}
		\caption{\textbf{Sampling error comparison with (green) and without (blue) cutoff and rounding error correction.} Shown are the distributions of the relative error of \textbf{A:} mean and \textbf{B:} variance.}
		\label{fig:sampling_err_corr}
	\end{figure}
	
\begin{figure}[!ht]
		\centering
		\includegraphics[width=\linewidth]{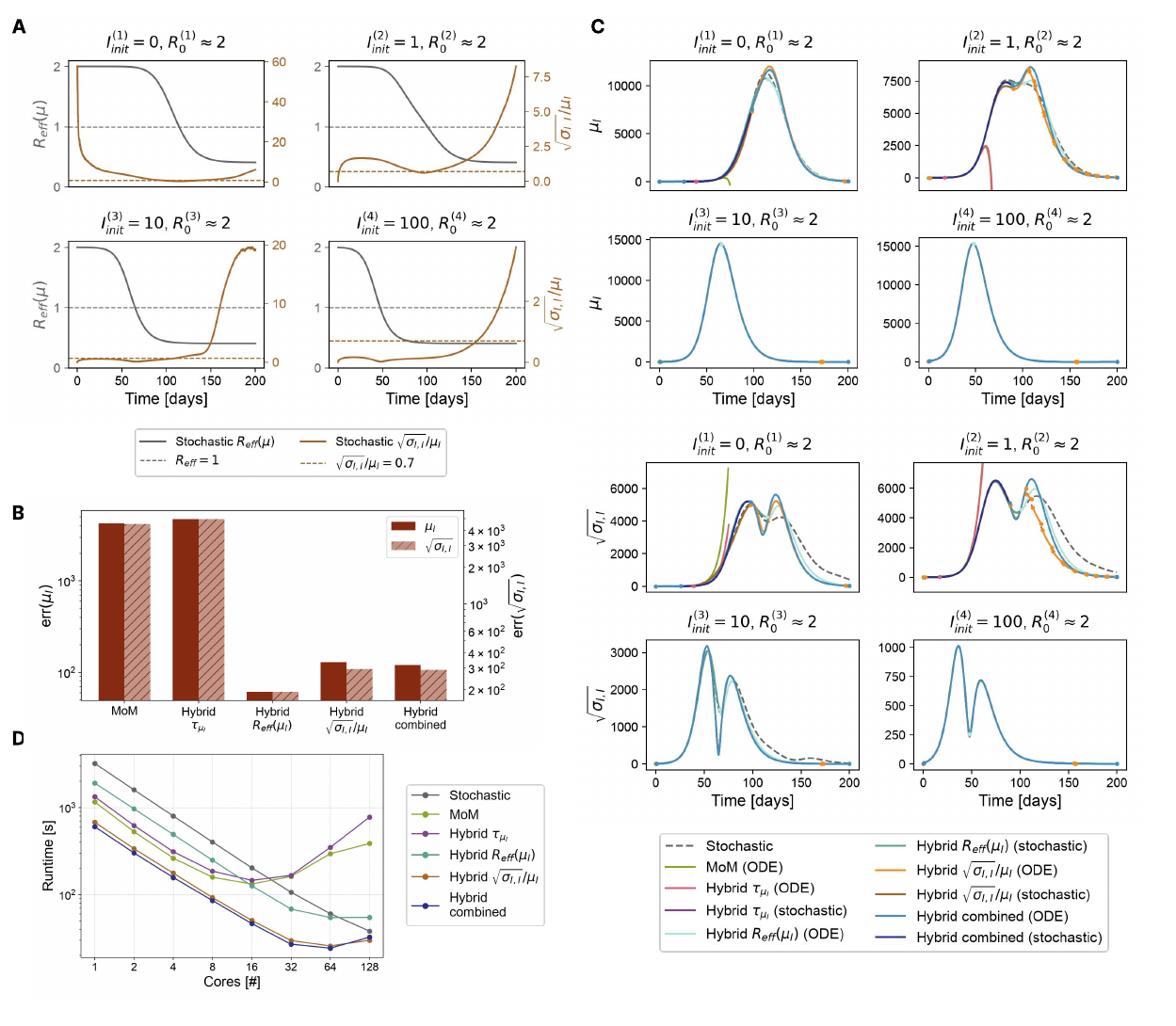}
		\caption{\textbf{Results of different condition functions for the SIR-type model with four regions, an initial reproduction number $R_0^{(k)}, k\in\{1,2,3,4\}$, of approximately two, and a varying number of initially infected per region.} For the stochastic model (and the stochastic parts of the hybrid models) $n_{sims}=10{,}000$ simulations are considered. \textbf{A:} Temporal evolution of the effective reproduction number of the mean (grey) and the relative standard deviation of the infected compartment (brown) for each region in the stochastic model. The horizontal dashed lines indicate the threshold values used for the corresponding conditions. \textbf{B:} Mean absolute error in mean and standard deviation of the moment-closure approximation and the hybrid model with four different conditions. \textbf{C:} Evolution of mean (top) and standard deviation (bottom) in every region for the stochastic model, the moment-closure approximation and the hybrid model with four different conditions. For the hybrid model the stochastic parts are displayed in a dark and the moment-closure approximation parts in a light color shade. The dot markers indicate switching time point in the hybrid models. \textbf{D:} Strong scaling of the stochastic model (Stochastic), the moment-closure approximation (MoM), and the hybrid models using the absolute threshold condition ($\tau_{\mu_I}$), the reproduction number condition ($R_{eff}(\mu_I)$), the standard deviation relation condition ($\frac{\sqrt{\sigma_{I,I}}}{\mu_I}$) and the combined condition.}
		\label{fig:hybrid_results_SIR_I0}
	\end{figure}

	To assess the potential of the hybrid modeling approach introduced in~\cref{sec:hybrid_modeling_approach}, we applied the spatio-temporal hybrid metapopulation model with the afore described conditions to both the SIR and SIRS model in a four-region setting with varying initial conditions in each region (\cref{fig:hybrid_results_SIR_I0,fig:hybrid_results_SIRS_I0}). We chose $\Delta t_{sw}=\Delta t_{ex}=1$, hence we evaluated the condition functions and exchanged individuals between models once per day. Applying the error correction~\cref{eq:sampling_dist} significantly reduces deviation from target statistics compared to sampling without error correction (\cref{fig:sampling_err_corr}).
	As in the spatio-temporal hybrid model some regions can be modeled stochastically while others are modeled with the MoM, the system states of both models were synchronized at the desired output time steps, so that the overall output time series is obtained by simply summing their individual time series. We therefore saved the system states of both the stochastic model and the moment-closure approximation in regular intervals of $0.1$ days. 
	Due to the heterogeneous initial infection levels, the temporal evolution of the mean, the reproduction number, and the relative standard deviation varies across regions. This leads to region-dependent hybrid switching time points (\cref{fig:hybrid_results_SIR_I0,fig:hybrid_results_SIRS_I0}A). Although region~$1$ starts without initially infected, an outbreak occurs due to spatial transitions of infected individuals from neighboring regions. Nevertheless, the outbreak timing and magnitude differ between regions (\cref{fig:hybrid_results_SIR_I0,fig:hybrid_results_SIRS_I0}C), highlighting the need for spatial resolution when predicting realistic epidemic dynamics.
	
	In regions $1$ and $2$, the moment-closure approximation and the hybrid model using the mean threshold condition $\tau_{\mu_{\iota(I,k)}}=10$ fail to capture the stochastic model's reference dynamics (\cref{fig:hybrid_results_SIR_I0,fig:hybrid_results_SIRS_I0}C). In contrast, the hybrid models using the standard deviation relation condition, the reproduction number condition and the combined condition are able to reproduce reference mean and variance. However, all hybrid models exhibit larger deviations from the stochastic model's mean in regions $1$ and $2$, where initial numbers of infected are small and stochastic fluctuations are more pronounced. The approximation of the standard deviation is less accurate than for the mean. Nevertheless, the hybrid model with the three aforementioned conditions is able to capture the overall temporal dynamics of it, including rise and decrease of the system's stochasticity. The hybrid model with the reproduction number condition performs best in terms of accuracy across all regions (\cref{fig:hybrid_results_SIR_I0,fig:hybrid_results_SIRS_I0}B), as it switches to the ODE-based formulation later in time than the other hybrid models. This in turn results in reduced runtime acceleration, however still achieving a speedup of $\approx40~\%$ on a single core (\cref{fig:hybrid_results_SIR_I0,fig:hybrid_results_SIRS_I0}D).
	
	Using OpenMP for shared-memory parallelization, we investigated the strong scaling behavior of all implementations. While for the stochastic model only straightforward parallelization across individual simulations was applied, the hybrid model and moment-closure approximation also employ intra-model parallelization. We observe almost ideal strong scaling for the stochastic model on up to 128 cores. In contrast, both the moment-closure approximation and the hybrid models saturate at 32 cores the latest. This saturation is not further investigated but may arise from a combination of factors such as insufficient workload per thread at higher core counts or a transition from compute-bound to memory-bound behavior. The hybrid model using the standard deviation relation condition is $4-5$ times faster for the SIR model (\cref{fig:hybrid_results_SIR_I0}D) and even $15-18$ times for the SIRS model (\cref{fig:hybrid_results_SIRS_I0}D) on up to 32 cores. Beyond the saturation point, runtimes of moment-closure approximations and hybrid models increase due to parallelization overhead. The fastest setup, which is the hybrid model using the combined condition on 32 cores, reduces the runtime of the stochastic model on 128 cores by $30$ and $80~\%$, respectively, while consuming only $25~\%$ of the computational resources.
	
	In a complementary scenario with identical numbers of initially infected across regions and varying region-dependent transmission rates, the hybrid model using the reproduction number condition again outperforms the other hybrid models and the moment-closure approximation in terms of accuracy, however achieving less runtime reduction (\cref{fig:supp_hybrid_results_SIR_R0,fig:supp_hybrid_results_SIRS_R0}).
	
	The results presented above emphasize that the choice of appropriate condition functions critically influences both the accuracy and computational performance of the hybrid model.
	
	\subsection{Modeling the collective dynamics of influenza-like illnesses in Germany}
	To demonstrate the applicability of the hybrid modeling approach for epidemic spread, we used it to study and simulate the dynamics of the collection of ILI in Germany from 1st August 2016 until 27th July 2021. In a first step, we considered a global model for Germany with only one region and used the combined condition described in the previous section as condition function; with a mean infected threshold of $100$ per $100{,}000$ inhabitants, a standard deviation of $0.3\mu_I$ in the infected compartment and a reproduction number threshold of $1$. Due to computational constraints, we modeled a population of size $835{,}771$, representing one hundredth of Germany's total population. We assumed an average time from infection until recovery of $7$ days; based on estimates for ILI symptom and shedding duration~\cite{morris_influenza_2024, ip_viral_2017}, yielding a recovery rate $\gamma=\frac{1}{7}$. We used the size of the infected compartment (prevalence) as an approximation for the reported incidence. Initial conditions for the mean number of infected $\mu_{I}(t_0)$ and recovered $\mu_{R}(t_0)$ as well as the immunity loss rate and the transmission rate function parameters were calibrated against reported estimates for 7-day ILI incidence~\cite{Buchholz_GrippeWeb_Daten_2026} from 1st August 2016 to 9th March 2020 (\cref{fig:supp_influenza_fitting}). Calibration was performed using approximate Bayesian computation with the sequential Monte Carlo algorithm implemented in pyABC~\cite{schalte_pyabc_2022}. Priors for the peak transmission rate $\lambda_0$ were scaled with the modeled population size. The final parameters can be found in~\cref{tab:influenza_parameters}.
	
	As reaction to the first confirmed case of SARS-CoV-2 in Germany on 25th February 2020, a series of NPIs were implemented starting from 16th March 2020 onward~\cite{a3m}, which also significantly altered transmission dynamics of ILI. To assess the impact of these interventions, we conducted a simulation study with the temporal-hybrid model and considered three scenarios:
	\begin{itemize}
		\item Baseline scenario (Base): Transmission dynamics continue unchanged from 9th March 2020 using season-specific parameters for the 2020/2021 ILI season (season $5$) set as the mean values across prior seasons.
		\item Global NPIs scenario: Multiplicative damping factors $d^{(1)},\ldots,d^{(6)}\in\mathbb{R}$ are applied to the transmission rate at six time points derived from~\cite{a3m}: $t_{d^{(1)}}=1323$ (16th March 2020), $t_{d^{(2)}}=1353$ (15th April 2020), $t_{d^{(3)}}=1554$ (2nd November 2020), $t_{d^{(4)}}=1577$ (25th November 2025), $t_{d^{(5)}}=1680$ (8th March 2021) and $t_{d^{(6)}}=1736$ (3rd May 2021). In this scenario $d^{(6)}=1$, indicating that the previous change applied at 8th March 2021 remains in effect, implying partial persistence of NPIs and sustained suppression of transmission.
		\item Global NPIs total lifting scenario: Multiplicative damping factors are applied as in the Global NPIs scenario, but all interventions are fully lifted on 3rd May 2021, and the peak transmission rate is set to the value from before 16th March 2020.
	\end{itemize}
	
\begin{figure}[!ht]
		\centering
		\includegraphics[width=\linewidth]{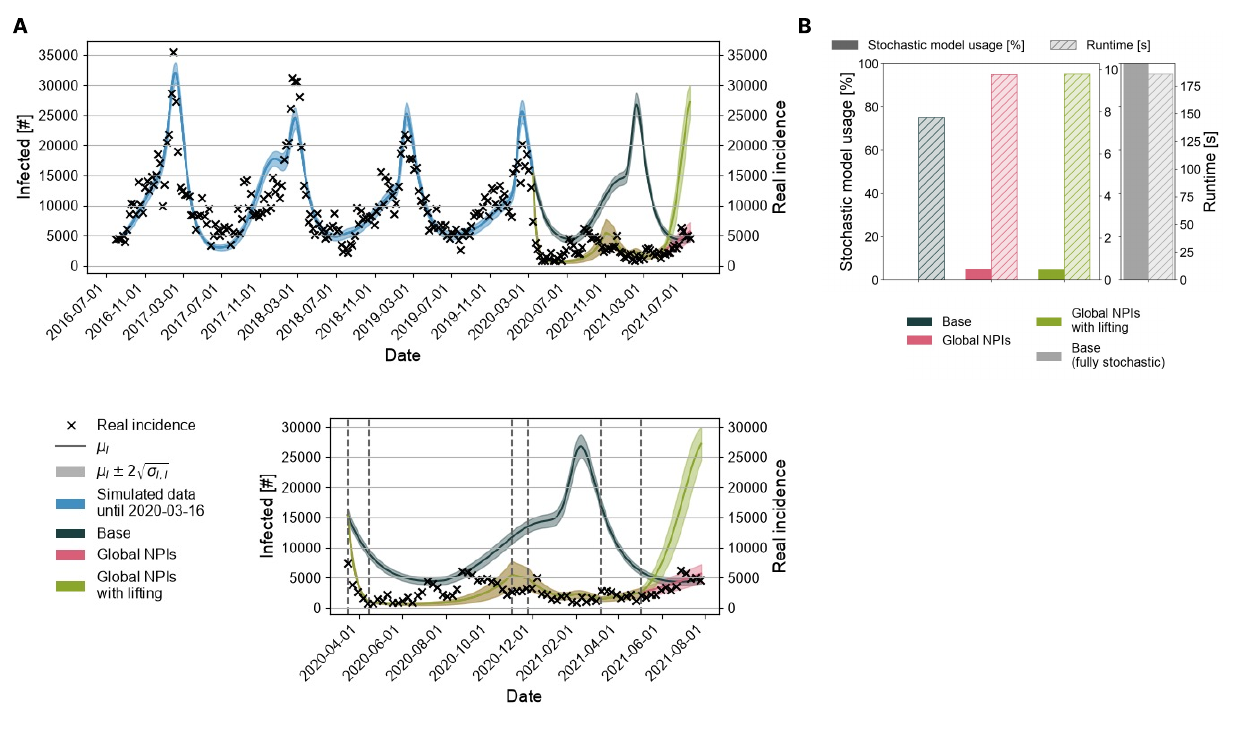}
		\caption{\textbf{Simulation outcomes of the hybrid model in three different scenarios for the evolution of ILI incidence in Germany.} \textbf{A:} Simulated prevalence trajectories of the hybrid model compared to real ILI incidence. The shown incidence corresponds to the reported weekly number of new infections scaled to match the modeled population size. Shown is the simulated mean number of infected (solid line) and twice the standard deviation (shaded area). The scenarios only differ from 16th March 2020 onwards when either no transmission rate dampings (Base), five dampings of varying strength (Global NPIs) or five dampings plus a sixth one that fully lifts the previous dampings from 3rd May 2021 onwards (Global NPIs total lifting) are applied. \textbf{B:} Runtime and relative number of time steps modeled with the stochastic formulation in the temporal hybrid model for all three scenarios compared to the Base scenario modeled fully stochastically.}
		\label{fig:influenza_germany}
	\end{figure}

	The reported incidence shows a significant reduction from 9th March 2020 onward compared to previous seasons (\cref{fig:influenza_germany}A). The baseline scenario fails to reproduce this reduction in incidence levels, instead predicting an epidemic wave similar to the previous seasons. In contrast, the global intervention setting including both scenarios (with and without lifting in May 2021) captures the observed decline and produces only a minor wave in November 2020. While the continued intervention scenario also aligns with the data after May 2021, the total lifting scenario predicts a substantial increase in infections in May-July 2021, which is inconsistent with the observed data.
	
	In the baseline scenario only the moment-closure approximation is used, while the other two scenarios switch to the stochastic model for about $5~\%$ of simulated time steps. Consequently, this leads to a runtime increase of approximately $30~\%$ compared to the baseline scenario (\cref{fig:influenza_germany}B). However, the two intervention scenarios are still $95~\%$ faster than simulating the baseline scenario with the stochastic model, which reflects the ability of the hybrid model to use the stochastic formulation only in critical regimes, while maintaining computational feasibility.
	
\begin{figure}[!ht]
		\centering
		\includegraphics[width=\linewidth]{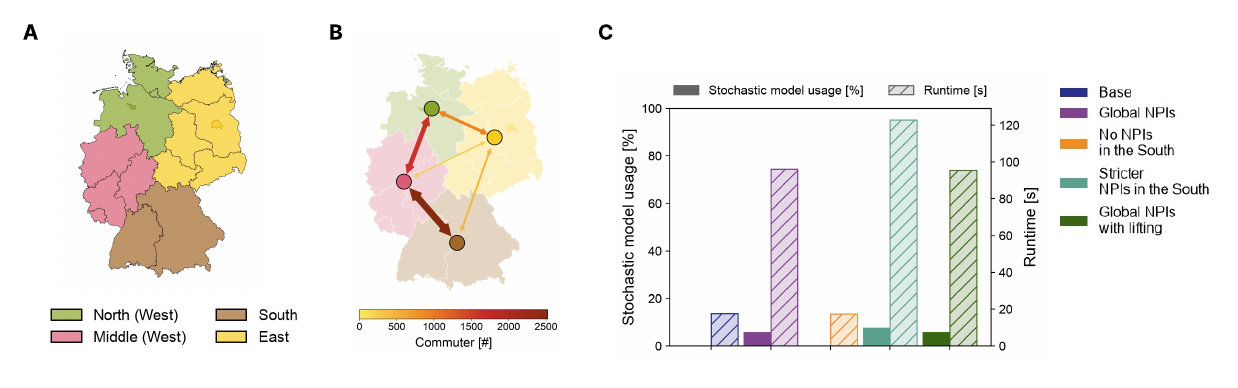}
		\caption{\textbf{Simulation outcomes of the hybrid model in five different scenarios for the spatially resolved evolution of ILI incidence in Germany.} \textbf{A:} Mapping of German federal states to modeled regions. \textbf{B:} Visualization of spatial transition between regions. \textbf{C:} Runtime and relative number of time steps modeled with the stochastic formulation in the spatio-temporal hybrid model for all five scenarios.} 
		\label{fig:influenza_germany_regional}
	\end{figure}

	In a second step, we extended the temporal-hybrid model for ILI to a spatio-temporal metapopulation model comprising four regions (\cref{fig:influenza_germany_regional}A), consistent with the spatially resolved data reported by~\cite{Buchholz_GrippeWeb_Daten_2026}. Unlike the global model, this regional model incorporates region-dependent transmission rates, enabling regionally heterogeneous NPIs. Furthermore, it integrates mobility patterns via spatial transition rates that were derived from a trip data set comprising activity-dependent daily trips between $400$ German counties~\cite{kuhn_vorlaufige_2023}. Based on~\cite{zunker_novel_2024}, we considered only work-related trips. The spatial transition rate $\kappa_z^{(k,l)}$, $z\in\{S,I,R\}$, from region $k$ to region $l$, $k\neq l$, is given by
	\begin{align*}
		\kappa_z^{(k,l)} = \frac{n_{trips}^{(k,l)}+n_{trips}^{(l,k)}}{N^{(k)}},
	\end{align*}
	with $n_{trips}^{(k,l)}, n_{trips}^{(l,k)}$ the daily number of work-related trips from region $k$ to region $l$ and vice versa; and $N^{(k)}$ the population size of region $k$. The spatial transition rates govern movement of individuals between regions (\cref{fig:influenza_germany_regional}B) and thus mediate inter-region transmission.
	
	In addition to the three scenarios defined for the global model (Base, Global NPIs, Global NPIs total lifting), we considered two supplementary scenarios that explicitly account for regional disparities in interventions. In these settings, we calibrated the regional transmission rates and initial mean values such that infection dynamics until 16th March 2020 match the spatially-resolved reported incidence data. We specifically considered the southern region and either implemented no intervention in this region or implemented a 10~\% stricter intervention set than in the other regions. More precisely, we considered the following two settings.
	\begin{itemize}
		\item No NPIs in the South: The global damping factors $d^{(1)},\ldots,d^{(6)}\in\mathbb{R}$ are applied to regions $1,2$ and $4$, while region~$3$ (South) remains unaffected by interventions and transmission dynamics in that region continue unchanged from 9th March 2020.
		\item Stricter NPIs in the South: The global damping factors are applied in region~$1,2$ and $4$, while in region~$3$ $10~\%$ stricter dampings are applied during the intervals $[t_{d^{(1)}},t_{d^{(2)}})$, $[t_{d^{(3)}},t_{d^{(4)}})$ and $[t_{d^{(4)}},t_{d^{(5)}})$.
	\end{itemize}
	
\begin{figure}[!ht]
		\centering
		\includegraphics[width=0.95\linewidth]{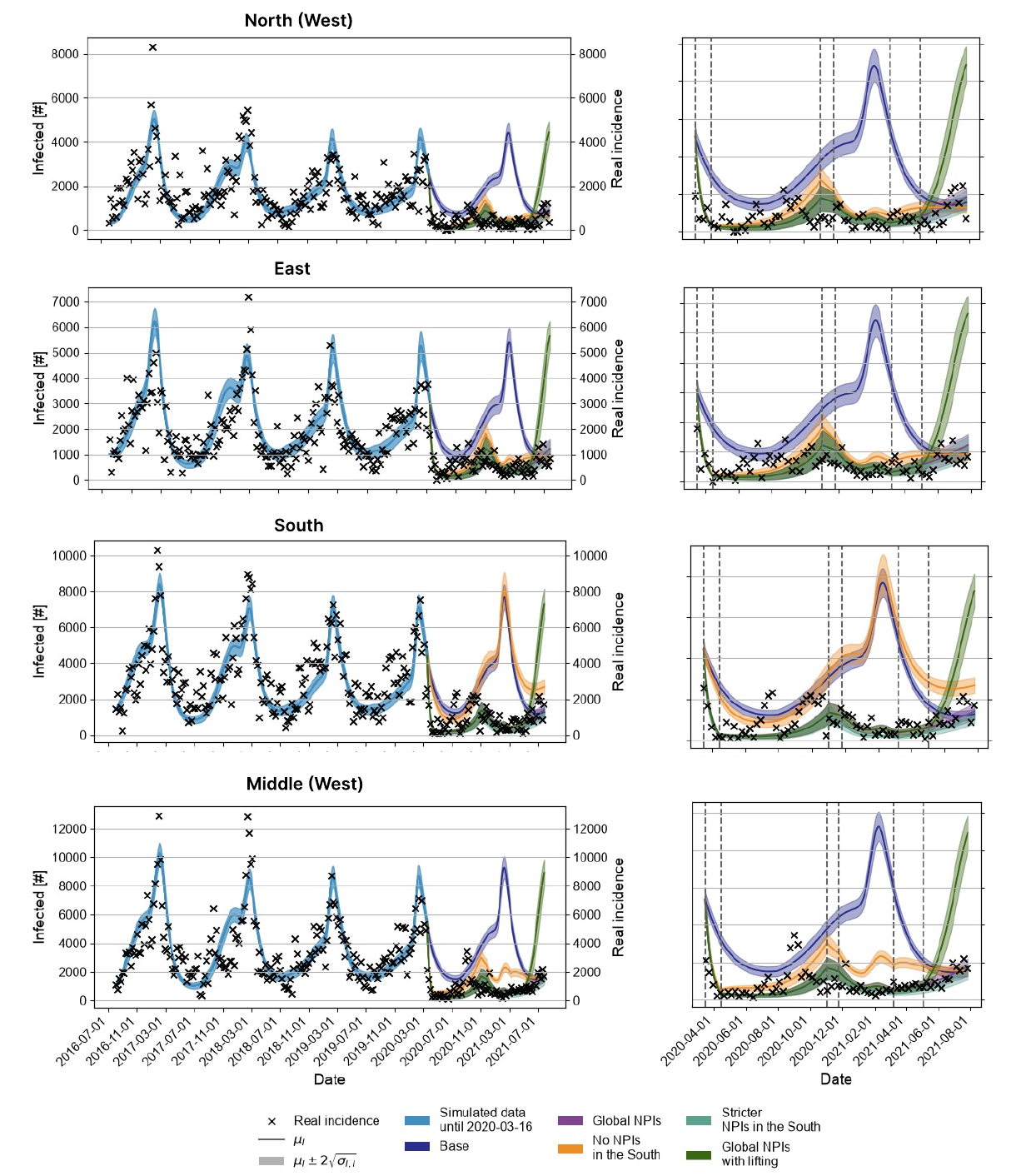}
		\caption{\textbf{Simulation outcomes of the hybrid model in five different scenarios for the spatially resolved evolution of ILI incidence in Germany.} From 16th March 2020 onward no dampings (Base), five (Global NPIs) or six (Global NPIs with lifting) identical dampings in all four regions, regionally heterogeneous dampings (Stricter NPIs in the South) or regionally heterogeneous lifting of dampings (No NPIs in the South) are applied. Shown is the whole modeled time frame (1st August 2016 - 27th July 2021) for each region and a zoomed in frame from the potential first damping time point until simulation end.} 
		\label{fig:influenza_germany_regional_trajectories}
	\end{figure}

	As in the global model, the runtime of the regional model depends on the number of time steps modeled stochastically. Scenarios with a higher proportion of time steps modeled stochastically exhibit correspondingly longer runtimes (\cref{fig:influenza_germany_regional}B).\\
	Incidence trajectories in the four regions reveal regionally heterogeneous patterns in ILI dynamics (\cref{fig:influenza_germany_regional_trajectories}). The baseline, global NPIs and global NPIs with total lifting scenarios produce similar trajectories to those observed in the global model. Among these, the global NPIs scenario without full lifting aligns best with reported incidence estimates. However, quantitative differences in peak prevalence and cumulative infections can be observed between regions, with region Middle (West) exhibiting the highest overall prevalence level. This may be attributed to its higher population density, resulting in increased contact rates, thereby enhancing transmission risk.
	
	The local scenarios (No NPIs in the South, Stricter NPIs in the South) reveal that regional intervention policies have non-local effects. In the no local interventions scenario, region~$3$ experiences an epidemic wave similar to the baseline scenario. While we only see slightly increased infection levels in the  North-Western and Eastern regions, we observe a relatively large number of infections in the Middle-Western region although interventions are in place. More precisely, besides region~$3$, local absence of interventions has the most pronounced effect in region~$4$ which experiences a second small infection wave from February to April 2021, following the wave in November 2020. This behavior is driven by the relatively high number of spatial transitions between region~$3$ and region~$4$, yielding sustained import of cases. In contrast, the scenario with $10~\%$ stricter NPIs in region~$3$ results in only marginal reductions in infection prevalence across all regions.
	
	\section{Discussion}
	In this study, we developed a spatio-temporal hybrid framework that allows each region of an epidemiological metapopulation model to switch independently between stochastic ensemble simulations and moment equations. The central methodological contribution is the use of ensemble-level statistics to guide local model choice. By retaining means, variances, and covariances during deterministic simulation, the framework accounts for epidemic variability beyond mean dynamics.
	
	First, we quantified the computational cost of stochastic ensembles and the moment-closure approximation by examining their scaling behavior in three different settings. This analysis revealed that as the number of regions increases, the runtime advantage of the moment-closure approximation over stochastic simulations diminishes. Conversely, larger total population sizes enhance the computational advantage of the moment-closure approximation compared to stochastic simulations. We systematically assessed the accuracy of the moment-closure approximation and used these results to formulate and compare switching criteria based on mean infection levels, relative variability, and the effective reproduction number. Finally, the application of the spatio-temporal-hybrid model to seasonal ILI demonstrates how regional interventions influence global dynamics and highlights the framework's potential for real-world epidemic forecasting. The seasonality function as well as applied dampings may lead to discontinuous transmission rates. Since dampings and season changes occurred only at full days, these discontinuities coincided with step boundaries and did therefore not cause integration problems. An alternative possible solution would be using sufficiently smooth interpolating transitions as discussed in~\cite{wendler_high_order}. Due to computational constraints, we only considered one hundredth of Germany's total population size and further assumed the average number of contacts to be proportional to the population size in a region. This assumption leads to reduced fluctuations in compartment sizes when increasing the population size, e.g. to the full population of Germany, which in turn influences the output of (switching) condition functions. Alternatively, regional transmission rates could be scaled by local population sizes, ensuring a constant average number of contacts.
	
	The presented framework extends our previous temporal and spatial hybridization approaches~\cite{bicker_hybrid_2025} by combining both forms of adaptation and applying switching criteria to summary statistics of the stochastic ensemble. Previous hybrid approaches for infectious disease dynamics have combined stochastic and deterministic descriptions in time, space, or both~\cite{bobashev_hybrid_2007,hunter_hybrid_2020,bradhurst_hybrid_2015,kehrer_hybrid_2025,bostanci2025}. Our emphasis is on linking the choice of representation to an empirical assessment of moment-closure accuracy, rather than relying solely on compartment counts in an individual realization. This enables the switching decision to account for variability across possible epidemic outcomes. The comparison of alternative criteria also demonstrates that their suitability depends on the dynamical regime and the desired balance between accuracy and computational cost. The conditions considered in this study use the moment-closure approximation if the mean number of infected is smaller than one. This condition is fulfilled in complete extinction settings, but also in scenarios with high extinction probability and a few surviving infections. Although the latter is not particularly likely for large population sizes, to explicitly exclude such scenarios from switching conditions, the cumulative size of the infected compartment could be used to inform switching conditions.
	
	Our approach further differs from hybrid methods that retain a reduced master equation for selected components of the system~\cite{hasenauer_method_2014,menz_hybrid_2012,jahnke_reduced_2011}. Such methods can be effective when the retained stochastic state space is sufficiently small, but this space may remain large for epidemiological compartments. Moment equations reduce computational effort in regimes where the closure provides an adequate approximation. Conversion and exchange rules connect the two representations, although reconstructing ensembles from a finite set of moments and approximating dependencies across their interface introduce additional errors that must be considered alongside moment-closure error. Future work should incorporate covariance contributions from incoming subpopulations and quantify the additional noise introduced by sampling variances uniformly within defined boundaries. Furthermore, extensions to non-polynomial propensities should be implemented (e.g., based on~\cite{KazeroonianThe2014}) to support common infection rates.
	
	We also leave for future work to extend the SIR- and SIRS-type models by an incidence compartment describing the flow from susceptible to infected. Such a compartment would circumvent using prevalence as a proxy for reported incidence, thereby facilitating the incorporation of and comparison with real-world data.
	
	Concluding, our study highlights the potential of hybrid stochastic-deterministic models to balance the trade-off between accuracy and computational complexity. While we demonstrated the applicability and performance of our framework in synthetic settings as well as for real-world scenario analysis, applying it to computationally intensive parameter estimation tasks offers further potential to reduce runtime and energy consumption.

	\section*{Acknowledgments}
	
	This work was supported by the Initiative and Networking Fund of the Helmholtz Association (grant agreement number KA1-Co-08, Project LOKI-Pandemics). It was furthermore supported by the German Federal Ministry of Education and Research and the German Federal Ministry of Research, Technology and Space under grant agreement 031L0325A (Project TwinChain). TwinChain is part of the Modeling Network for Severe Infectious Diseases (MONID). Additionally, this work was supported by the European Union via the ERC grant INTEGRATE, grant agreement number 101126146, and under Germany’s Excellence Strategy by the Deutsche Forschungsgemeinschaft (DFG, German Research Foundation) (EXC 2047—390685813, EXC 2151—390873048, and 524747443), and by the University of Bonn via the Schlegel Professorship of J.H. 
	
	\section*{Declaration of competing interest}
	The authors declare to not have any conflict of interest.
	
	\section*{Data availability}
	The developed models as well as all simulation code for this study are implemented within the MEmilio software framework and publicly available at \url{https://github.com/jubicker/memilio/tree/hybrid2}. Visualized simulation data can be shared upon request.
	
	\section*{CRediT authorship contribution statement}
	\noindent\textbf{Julia Bicker:} Writing – Review \& Editing, Writing – Original Draft, Conceptualization, Data Curation, Formal Analysis, Investigation, Methodology, Software, Validation, Visualization. \textbf{Kilian Volmer:} Writing – Review \& Editing, Conceptualization, Software, Validation. \textbf{Jan Hasenauer:} Writing – Review \& Editing, Conceptualization, Funding Acquisition, Project Administration, Supervision, Validation. \textbf{Martin J. Kühn:} Writing – Review \& Editing, Conceptualization, Funding Acquisition, Project Administration, Supervision, Validation.
	
	\bibliographystyle{elsarticle-num}
	\bibliography{literature.bib}

@inproceedings{KazeroonianThe2014,
	address = {Cape Town, South Africa},
	author = {Kazeroonian, A. and Theis, F. J. and Hasenauer, J.},
	booktitle = {Proceedings of the 19th {IFAC} World Congress},
	doi = {10.3182/20140824-6-ZA-1003.02298},
	month = {Aug.},
	number = {1},
	pages = {1729--1735},
	title = {Modeling of stochastic biological processes with non-polynomial propensities using non-central conditional moment equation},
	url = {http://dx.doi.org/10.3182/20140824-6-ZA-1003.02298},
	volume = {19},
	year = {2014}}

@article{ploetzke_lct_2025,
	title = {Revisiting the Linear Chain Trick in epidemiological models: Implications of underlying assumptions for numerical solutions},
	journal = {Mathematics and Computers in Simulation},
	volume = {239},
	pages = {823-844},
	year = {2026},
	issn = {0378-4754},
	doi = {https://doi.org/10.1016/j.matcom.2025.07.045},
	author = {Lena Plötzke and Anna Wendler and René Schmieding and Martin J. Kühn}
}

@article{bicker_hybrid_2025,
	title = {Hybrid metapopulation agent-based epidemiological models for efficient insight on the individual scale: A contribution to green computing},
	journal = {Infectious Disease Modelling},
	volume = {10},
	number = {2},
	pages = {571-590},
	year = {2025},
	issn = {2468-0427},
	doi = {https://doi.org/10.1016/j.idm.2024.12.015},
	author = {Julia Bicker and René Schmieding and Michael Meyer-Hermann and Martin J. Kühn}
}

@article{KERKMANN_ABM_2025,
	title = {Agent-based modeling for realistic reproduction of human mobility and contact behavior to evaluate test and isolation strategies in epidemic infectious disease spread},
	journal = {Computers in Biology and Medicine},
	volume = {193},
	pages = {110269},
	year = {2025},
	issn = {0010-4825},
	doi = {https://doi.org/10.1016/j.compbiomed.2025.110269},
	url = {https://doi.org/10.1016/j.compbiomed.2025.110269},
	author = {David Kerkmann and Sascha Korf and Khoa Nguyen and Daniel Abele and Alain Schengen and Carlotta Gerstein and Jens Henrik Göbbert and Achim Basermann and Martin J. Kühn and Michael Meyer-Hermann}
}

@article{kuhn_assessment_2021,
	title = {Assessment of effective mitigation and prediction of the spread of {SARS}-{CoV}-2 in {Germany} using demographic information and spatial resolution},
	issn = {0025-5564},
	url = {https://doi.org/10.1016/j.mbs.2021.108648},
	doi = {https://doi.org/10.1016/j.mbs.2021.108648},
	journal = {Mathematical Biosciences},
	author = {K\"uhn, Martin J. and Abele, Daniel and Mitra, Tanmay and Koslow, Wadim and Abedi, Majid and Rack, Kathrin and Siggel, Martin and Khailaie, Sahamoddin and Klitz, Margrit and Binder, Sebastian and Spataro, Luca and Gilg, Jonas and Kleinert, Jan and Häberle, Matthias and Plötzke, Lena and Spinner, Christoph D. and Stecher, Melanie and Zhu, Xiao Xiang and Basermann, Achim and Meyer-Hermann, Michael},
	year = {2021},
	pages = {108648}
}

@article{winkelmann_mathematical_2021,
	title = {Mathematical modeling of spatio-temporal population dynamics and application to epidemic spreading},
	volume = {336},
	issn = {0025-5564},
	doi = {https://doi.org/10.1016/j.mbs.2021.108619},
	journal = {Mathematical Biosciences},
	author = {Winkelmann, Stefanie and Zonker, Johannes and Schütte, Christof and Conrad, Nataša Djurdjevac},
	year = {2021},
	pages = {108619}
}

@article{kerr_covasim_2021,
	title = {Covasim: {An} agent-based model of {COVID}-19 dynamics and interventions},
	volume = {17},
	url = {https://doi.org/10.1371/journal.pcbi.1009149},
	doi = {10.1371/journal.pcbi.1009149},
	number = {7},
	journal = {PLOS Computational Biology},
	author = {Kerr, Cliff C. and Stuart, Robyn M. and Mistry, Dina and Abeysuriya, Romesh G. and Rosenfeld, Katherine and Hart, Gregory R. and Núñez, Rafael C. and Cohen, Jamie A. and Selvaraj, Prashanth and Hagedorn, Brittany and George, Lauren and Jastrzebski, Michal and Izzo, Amanda S. and Fowler, Greer and Palmer, Anna and Delport, Dominic and Scott, Nick and Kelly, Sherrie L. and Bennette, Caroline S. and Wagner, Bradley G. and Chang, Stewart T. and Oron, Assaf P. and Wenger, Edward A. and Panovska-Griffiths, Jasmina and Famulare, Michael and Klein, Daniel J.},
	month = jul,
	year = {2021},
	pages = {1--32},
}

@article{Ahnert_boost,
	author = {Ahnert, Karsten and Mulansky, Mario},
	title = {Odeint – {Solving Ordinary Differential Equations in C++}},
	journal = {AIP Conference Proceedings},
	volume = {1389},
	number = {1},
	pages = {1586-1589},
	year = {2011},
	month = {09},
	issn = {0094-243X},
	doi = {10.1063/1.3637934},
	url = {https://doi.org/10.1063/1.3637934}
}

@article{zunker_novel_2024,
	doi = {10.1371/journal.pcbi.1012630},
	url = {https://doi.org/10.1371/journal.pcbi.1012630},
	author = {Zunker, Henrik AND Schmieding, René AND Kerkmann, David AND Schengen, Alain AND Diexer, Sophie AND Mikolajczyk, Rafael AND Meyer-Hermann, Michael AND Kühn, Martin J.},
	journal = {PLOS Computational Biology},
	publisher = {Public Library of Science},
	title = {Novel travel time aware metapopulation models and multi-layer waning immunity for late-phase epidemic and endemic scenarios},
	year = {2024},
	month = {Dez},
	volume = {20},
	number = {12}
}

@article{bradhurst_hybrid_2015,
	title = {A hybrid modeling approach to simulating foot-and-mouth disease outbreaks in {Australian} livestock},
	volume = {3},
	issn = {2296-665X},
	doi = {10.3389/fenvs.2015.00017},
	urldate = {2023-02-06},
	journal = {Frontiers in Environmental Science},
	author = {Bradhurst, Richard A. and Roche, Sharon E. and East, Iain J. and Kwan, Paul and Garner, M. Graeme},
	month = mar,
	year = {2015}
}

@article{schalte_pyabc_2022,
	title = {{pyABC}: {Efficient} and robust easy-to-use approximate {Bayesian} computation},
	volume = {7},
	issn = {2475-9066},
	shorttitle = {{pyABC}},
	url = {https://doi.org/10.21105/joss.04304},
	doi = {10.21105/joss.04304},
	language = {en},
	number = {74},
	urldate = {2023-02-26},
	journal = {Journal of Open Source Software},
	author = {Schälte, Yannik and Klinger, Emmanuel and Alamoudi, Emad and Hasenauer, Jan},
	month = jun,
	year = {2022},
	pages = {4304}
}

@article{bostanci2025,
	title = {Integrating Agent-Based and Compartmental Models for Infectious Disease Modeling: A Novel Hybrid Approach},
	author = {Bostanci, Inan and Conrad, Tim},
	journal = {Journal of Artificial Societies and Social Simulation},
	ISSN = {1460-7425},
	volume = {28},
	number = {1},
	pages = {5},
	year = {2025},
	DOI = {10.18564/jasss.5567}
}

@techreport{ecdc_npis_2024,
	institution ={European Centre for Disease Prevention and Control},
	title = {Public health and social measures for health emergencies and pandemics in the {EU}/{EEA}: recommendations for strengthening preparedness planning},
	shorttitle = {Public health and social measures for health emergencies and pandemics in the {EU}/{EEA}},
	language = {en},
	urldate = {2024-09-29},
	author = {{European Centre for Disease Prevention and Control}},
	year = {2024},
	doi = {10.2900/253991}
}

@article{shattock_impact_2022,
	title = {Impact of vaccination and non-pharmaceutical interventions on {SARS}-{CoV}-2 dynamics in {Switzerland}},
	volume = {38},
	issn = {1755-4365},
	url = {https://doi.org/10.1016/j.epidem.2021.100535},
	doi = {10.1016/j.epidem.2021.100535},
	urldate = {2025-12-18},
	journal = {Epidemics},
	author = {Shattock, Andrew J. and Le Rutte, Epke A. and Dünner, Robert P. and Sen, Swapnoleena and Kelly, Sherrie L. and Chitnis, Nakul and Penny, Melissa A.},
	month = mar,
	year = {2022},
	pages = {100535},
}

@article{Edlund2011,
	title={Comparing three basic models for seasonal influenza},
	author={Edlund, Stefan and Kaufman, James and Lessler, Justin and Douglas, Judith and Bromberg, Michal and Kaufman, Zalman and Bassal, Ravit and Chodick, Gabriel and Marom, Rachel and Shalev, Varda and others},
	journal={Epidemics},
	volume={3},
	number={3-4},
	pages={135--142},
	year={2011},
	publisher={Elsevier},
	doi = {10.1016/j.epidem.2011.04.002}
}

@article{bicker2026memilio,
	title={MEmilio -- A high performance Modular EpideMIcs simuLatIOn software for multi-scale and comparative simulations of infectious disease dynamics}, 
	author={Julia Bicker and Carlotta Gerstein and David Kerkmann and Sascha Korf and René Schmieding and Anna Wendler and Henrik Zunker and Daniel Abele and Maximilian Betz and Khoa Nguyen and Lena Plötzke and Kilian Volmer and Agatha Schmidt and Nils Waßmuth and Patrick Lenz and Daniel Richter and Hannah Tritzschak and Ralf Hannemann-Tamas and Julian Litz and Paul Johannssen and Marielena Borges and Annika Jungklaus and Manuel Heger and Annalena Lange and Elisabeth Kluth and Kathrin Rack and Vincent Wieland and Jonas Arruda and Sebastian Binder and Margrit Klitz and Martin Siggel and Manuel Dahmen and Achim Basermann and Michael Meyer-Hermann and Jan Hasenauer and Martin J. Kühn},
	year={2026},
	journal={Scientific Reports},
    volume = {16},    
    number={1},
    pages={26950},    
	doi={10.1038/s41598-026-66481-6}
}

@article{gillespie_exact_1977,
	title = {Exact stochastic simulation of coupled chemical reactions},
	volume = {81},
	issn = {0022-3654},
	url = {https://doi.org/10.1021/j100540a008},
	doi = {10.1021/j100540a008},
	number = {25},
	journal = {The Journal of Physical Chemistry},
	author = {Gillespie, Daniel T.},
	month = dec,
	year = {1977},
	pages = {2340--2361}
}

@article{khailaie_development_2021,
	title = {Development of the reproduction number from coronavirus {SARS}-{CoV}-2 case data in {Germany} and implications for political measures},
	volume = {19},
	number = {1},
	journal = {BMC medicine},
	author = {Khailaie, Sahamoddin and Mitra, Tanmay and Bandyopadhyay, Arnab and Schips, Marta and Mascheroni, Pietro and Vanella, Patrizio and Lange, Berit and Binder, Sebastian C and Meyer-Hermann, Michael},
	year = {2021},
	pages = {1--16}
}

@article{hunter_hybrid_2020,
	title = {A Hybrid agent-based and Equation based Model for the Spread of Infectious Diseases},
	volume = {23},
	issn = {1460-7425},
doi = {10.18564/jasss.4421},
	number = {4},
	journal = {Journal of Artificial Societies and Social Simulation},
	author = {Hunter, Elizabeth and Mac Namee, Brian and Kelleher, John},
	year = {2020},
	pages = {14}
}

@inproceedings{kasereka_hybrid_2014,
	address = {Algiers, Algeria},
	title = {A hybrid model for modeling the spread of epidemics: {Theory} and simulation},
	isbn = {978-1-4799-7508-2 978-1-4799-7507-5},
	shorttitle = {A hybrid model for modeling the spread of epidemics},
	doi = {10.1109/ISKO-Maghreb.2014.7033457},
	language = {en},
	urldate = {2024-06-04},
	booktitle = {2014 4th {International} {Symposium} {ISKO}-{Maghreb}: {Concepts} and {Tools} for knowledge {Management} ({ISKO}-{Maghreb})},
	publisher = {IEEE},
	author = {Kasereka, Selain and Kasoro, Nathanael and Chokki, Abiola Paterne},
	month = nov,
	year = {2014},
	pages = {1--7}
}

@article{yoneyama_hybrid_2012,
	title = {A hybrid Model for Disease Spread and an Application to the {SARS} Pandemic},
	volume = {15},
	issn = {1460-7425},
doi = {10.18564/jasss.1782},
	language = {en},
	number = {1},
	urldate = {2024-06-04},
	journal = {Journal of Artificial Societies and Social Simulation},
	author = {Yoneyama, Teruhiko and Das, Sanmay and Krishnamoorthy, Mukkai},
	year = {2012},
	pages = {5}
}

@article{liu_modelling_2022,
	title = {Modelling effectiveness of {COVID}-19 pandemic control policies using an {Area}-based {SEIR} model with consideration of infection during interzonal travel},
	volume = {161},
	issn = {09658564},
doi = {10.1016/j.tra.2022.05.003},
	language = {en},
	urldate = {2023-06-26},
	journal = {Transportation Research Part A: Policy and Practice},
	author = {Liu, Jielun and Ong, Ghim Ping and Pang, Vincent Junxiong},
	month = jul,
	year = {2022},
	pages = {25--47}
}

@inproceedings{bobashev_hybrid_2007,
	address = {Washington, DC, USA},
	title = {A hybrid Epidemic Model: Combining The Advantages Of agent-based {And} {Equation}-based Approaches},
	isbn = {978-1-4244-1305-8},
	shorttitle = {A hybrid {Epidemic} {Model}},
	doi = {10.1109/WSC.2007.4419767},
	language = {en},
	urldate = {2023-10-04},
	booktitle = {2007 {Winter} {Simulation} {Conference}},
	publisher = {IEEE},
	author = {Bobashev, Georgiy V. and Goedecke, D. Michael and {Feng Yu} and Epstein, Joshua M.},
	year = {2007},
	pages = {1532--1537}
}

@software{memilio,
	author = {Bicker, Julia and Kerkmann, David and Korf, Sascha Alexander and Plötzke, Lena and Schmieding, René and Wendler, Anna Clara and Zunker, Henrik and Nguyen, Khoa and Abele, Daniel and Gerstein, Carlotta and Lenz, Patrick and Betz, Maximilian Franz and Schmidt, Agatha and Hannemann-Tamas, Ralf and Volmer, Kilian and Waßmuth, Nils and Tritzschak, Hannah and Richter, Daniel and Heger, Manuel and Basermann, Achim and Meyer-Hermann, Michael and Hasenauer, Jan and Kühn, Martin Joachim},
	title = {{MEmilio v2.3.0 - A high performance Modular EpideMIcs simuLatIOn software}},
	url = {https://doi.org/10.5281/zenodo.18929098},
	version = {2.3.0},
	year = {2026}
}

@article{rozhnova_model-based_2021,
	title = {Model-based evaluation of school- and non-school-related measures to control the {COVID}-19 pandemic},
	volume = {12},
	doi = {10.1038/s41467-021-21899-6},
	number = {1},
	journal = {Nature Communications},
	author = {Rozhnova, Ganna and Van Dorp, Christiaan H. and Bruijning-Verhagen, Patricia and Bootsma, Martin C. J. and Van De Wijgert, Janneke H. H. M. and Bonten, Marc J. M. and Kretzschmar, Mirjam E.},
	month = Mar,
	year = {2021}
}

@article{hurtado_generalizations_2019,
	title = {Generalizations of the ‘{Linear} {Chain} {Trick}’: incorporating more flexible dwell time distributions into mean field {ODE} models},
	doi = {10.1007/s00285-019-01412-w},
	journal = {Journal of Mathematical Biology},
	author = {Hurtado, Paul J. and Kirosingh, Adam S.},
	month = oct,
	year = {2019},
	number={5},
	volume={79}
}

@article{kehrer_hybrid_2025,
	title = {Hybrid {PDE}–{ODE} models for efficient simulation of infection spread in epidemiology},
	volume = {481},
	issn = {1364-5021},
	doi = {10.1098/rspa.2024.0421},
	number = {2305},
	urldate = {2025-12-21},
	journal = {Proceedings of the Royal Society A: Mathematical, Physical and Engineering Sciences},
	author = {Kehrer, Kristina and Weiser, Martin and Conrad, Tim},
	month = jan,
	year = {2025},
	pages = {20240421}
}

@article{chao_flute_2010,
	title = {{FluTE}, a Publicly Available Stochastic Influenza Epidemic Simulation Model},
	volume = {6},
	issn = {1553-7358},
doi = {10.1371/journal.pcbi.1000656},
	pages = {e1000656},
	number = {1},
	journal = {{PLOS} Computational Biology},
	shortjournal = {{PLOS} Computational Biology},
	publisher = {Public Library of Science},
	author = {Chao, Dennis L. and Halloran, M. Elizabeth and Obenchain, Valerie J. and Jr, Ira M. Longini},
	urldate = {2025-01-15},
	date = {2010-01-29},
	langid = {english}
}

@article{giordano_modelling_2020,
	title = {Modelling the {COVID}-19 epidemic and implementation of population-wide interventions in {Italy}},
	volume = {26},
	issn = {1078-8956, 1546-170X},
	doi = {10.1038/s41591-020-0883-7},
	language = {en},
	number = {6},
	urldate = {2023-09-18},
	journal = {Nature Medicine},
	author = {Giordano, Giulia and Blanchini, Franco and Bruno, Raffaele and Colaneri, Patrizio and Di Filippo, Alessandro and Di Matteo, Angela and Colaneri, Marta},
	month = jun,
	year = {2020},
	pages = {855--860}
}

@misc{population_statistic,
	author       = {{Statistische Ämter des Bundes und der Länder}},
	title        = {{Bevölkerung nach Geschlecht und Altersgruppen (17) – Stichtag 31.12. – regionale Tiefe: Gemeinden}},
	howpublished = {\url{https://www.regionalstatistik.de/genesis/online?operation=table&code=12411-02-03-5}},
	note         = {Accessed: 2025-04-22},
	year         = {2025}
}

@misc{kuhn_vorlaufige_2023,
	title = {Vorläufige bundesweite {Verkehrsströme}},
	url = {https://mobilithek.info/offers/573360269906817024},
	urldate = {2024-03-01},
	author = {Kühn, Martin and Schengen, Alain and Mocanu, Tudor and Joshi, Jigeeshu},
	month = mar,
	year = {2023}
}

@article{ArinoVdD2003a,
	title = {A multi-city epidemic model},
	volume = {10},
	ISSN = {1547-724X},
	DOI = {10.1080/08898480306720},
	number = {3},
	journal = {Mathematical Population Studies},
	publisher = {Informa UK Limited},
	author = {Arino,  Julien and van den Driessche,  P.},
	year = {2003},
	month = jan,
	pages = {175–193}
}

@article{
Citron2021,
	author = {Daniel T. Citron  and Carlos A. Guerra  and Andrew J. Dolgert  and Sean L. Wu  and John M. Henry  and Héctor M. Sánchez C.  and David L. Smith },
	title = {Comparing metapopulation dynamics of infectious diseases under different models of human movement},
	journal = {Proceedings of the National Academy of Sciences},
	volume = {118},
	number = {18},
	pages = {e2007488118},
	year = {2021},
	doi = {10.1073/pnas.2007488118}}

@article{COSNER2009550,
	title = {The effects of human movement on the persistence of vector-borne diseases},
	journal = {Journal of Theoretical Biology},
	volume = {258},
	number = {4},
	pages = {550-560},
	year = {2009},
	issn = {0022-5193},
	doi = {https://doi.org/10.1016/j.jtbi.2009.02.016},
	author = {C. Cosner and J.C. Beier and R.S. Cantrell and D. Impoinvil and L. Kapitanski and M.D. Potts and A. Troyo and S. Ruan}
}

@inproceedings{wang2018,
	author = {Wang, Jingyuan and Wang, Xiaojian and Wu, Junjie},
	title = {Inferring Metapopulation Propagation Network for Intra-city Epidemic Control and Prevention},
	year = {2018},
	isbn = {9781450355520},
	publisher = {Association for Computing Machinery},
	address = {New York, NY, USA},
	doi = {10.1145/3219819.3219865},
	booktitle = {Proceedings of the 24th ACM SIGKDD International Conference on Knowledge Discovery \& Data Mining},
	pages = {830–838},
	numpages = {9},
	location = {London, United Kingdom},
	series = {KDD '18}
}

@article{lee_moment_2009,
	title = {A moment closure method for stochastic reaction networks},
	volume = {130},
	issn = {0021-9606, 1089-7690},
	doi = {10.1063/1.3103264},
	language = {en},
	number = {13},
	urldate = {2025-08-04},
	journal = {The Journal of Chemical Physics},
	author = {Lee, Chang Hyeong and Kim, Kyeong-Hun and Kim, Pilwon},
	month = apr,
	year = {2009},
	pages = {134107},
}

@article{achterberg_sis_2022,
  title = {Moment closure approximations of susceptible-infected-susceptible epidemics on adaptive networks},
  author = {Achterberg, Massimo A. and Van Mieghem, Piet},
  journal = {Phys. Rev. E},
  volume = {106},
  issue = {1},
  pages = {014308},
  numpages = {21},
  year = {2022},
  month = {Jul},
  publisher = {American Physical Society},
  doi = {10.1103/PhysRevE.106.014308},
}

@article{krishnarajah_novel_2005,
	title = {Novel moment closure approximations in stochastic epidemics},
	volume = {67},
	issn = {00928240},
	doi = {10.1016/j.bulm.2004.11.002},
	language = {en},
	number = {4},
	urldate = {2026-08-05},
	journal = {Bulletin of Mathematical Biology},
	author = {Krishnarajah, I and Cook, A and Marion, G and Gibson, G},
	month = jul,
	year = {2005},
	pages = {855--873},
}

@article{engblomComputingmoments2006,
	title = {Computing the moments of high dimensional solutions of the master equation},
	volume = {180},
	issn = {00963003},
	doi = {10.1016/j.amc.2005.12.032},
	language = {en},
	number = {2},
	urldate = {2023-02-22},
	journal = {Applied Mathematics and Computation},
	author = {Engblom, Stefan},
	month = sep,
	year = {2006},
	pages = {498--515},
}

@article{hasenauer_method_2014,
	title = {Method of conditional moments ({MCM}) for the {Chemical} {Master} {Equation}: {A} unified framework for the method of moments and hybrid stochastic-deterministic models},
	volume = {69},
	copyright = {http://www.springer.com/tdm},
	issn = {0303-6812, 1432-1416},
	shorttitle = {Method of conditional moments ({MCM}) for the {Chemical} {Master} {Equation}},
	doi = {10.1007/s00285-013-0711-5},
	language = {en},
	number = {3},
	urldate = {2026-08-11},
	journal = {Journal of Mathematical Biology},
	author = {Hasenauer, J. and Wolf, V. and Kazeroonian, A. and Theis, F. J.},
	month = sep,
	year = {2014},
	pages = {687--735},
}

@inproceedings{singh_lognormal_2006,
	address = {San Diego, CA, USA},
	title = {Lognormal {Moment} {Closures} for {Biochemical} {Reactions}},
	isbn = {978-1-4244-0171-0},
	doi = {10.1109/CDC.2006.376994},
	urldate = {2026-08-13},
	booktitle = {Proceedings of the 45th {IEEE} {Conference} on {Decision} and {Control}},
	publisher = {IEEE},
	author = {Singh, Abhyudai and Hespanha, Joao Pedro},
	year = {2006},
	pages = {2063--2068},
}

@article{bronstein_variational_2018,
	title = {A variational approach to moment-closure approximations for the kinetics of biomolecular reaction networks},
	volume = {148},
	issn = {0021-9606, 1089-7690},
	doi = {10.1063/1.5003892},
	language = {en},
	number = {1},
	urldate = {2026-08-13},
	journal = {The Journal of Chemical Physics},
	author = {Bronstein, Leo and Koeppl, Heinz},
	month = jan,
	year = {2018},
	pages = {014105}
}

@article{singh_derivative_2007,
	title = {A {Derivative} {Matching} {Approach} to {Moment} {Closure} for the {Stochastic} {Logistic} {Model}},
	volume = {69},
	copyright = {http://www.springer.com/tdm},
	issn = {0092-8240, 1522-9602},
	doi = {10.1007/s11538-007-9198-9},
	language = {en},
	number = {6},
	urldate = {2026-08-13},
	journal = {Bulletin of Mathematical Biology},
	author = {Singh, Abhyudai and Hespanha, João Pedro},
	month = aug,
	year = {2007},
	pages = {1909--1925},
}

@article{singh_approximate_2011,
	title = {Approximate {Moment} {Dynamics} for {Chemically} {Reacting} {Systems}},
	volume = {56},
	copyright = {https://ieeexplore.ieee.org/Xplorehelp/downloads/license-information/IEEE.html},
	issn = {0018-9286},
	doi = {10.1109/TAC.2010.2088631},
	number = {2},
	urldate = {2026-08-13},
	journal = {IEEE Transactions on Automatic Control},
	author = {Singh, Abhyudai and Hespanha, João P.},
	month = feb,
	year = {2011},
	pages = {414--418},
}

@article{petrova_evolution_2018,
	title = {The evolution of seasonal influenza viruses},
	volume = {16},
	issn = {1740-1526, 1740-1534},
	doi = {10.1038/nrmicro.2017.118},
	language = {en},
	number = {1},
	urldate = {2026-08-13},
	journal = {Nature Reviews Microbiology},
	author = {Petrova, Velislava N. and Russell, Colin A.},
	month = jan,
	year = {2018},
	pages = {47--60},
}

@article{bull_waning_2025,
	title = {Waning immunity drives respiratory virus evolution and reinfection},
	volume = {13},
	copyright = {https://creativecommons.org/licenses/by/4.0/},
	issn = {2050-6201},
	doi = {10.1093/emph/eoaf002},
	language = {en},
	number = {1},
	urldate = {2026-08-13},
	journal = {Evolution, Medicine, and Public Health},
	author = {Bull, James J and Koelle, Katia and Antia, Rustom},
	month = jan,
	year = {2025},
	pages = {101--110},
}

@article{moriyama_seasonality_2020,
	title = {Seasonality of {Respiratory} {Viral} {Infections}},
	volume = {7},
	issn = {2327-056X, 2327-0578},
	doi = {10.1146/annurev-virology-012420-022445},
	language = {en},
	number = {1},
	urldate = {2026-08-13},
	journal = {Annual Review of Virology},
	author = {Moriyama, Miyu and Hugentobler, Walter J. and Iwasaki, Akiko},
	month = sep,
	year = {2020},
	pages = {83--101},
}

@article{morris_influenza_2024,
	title = {Influenza virus shedding and symptoms: {Dynamics} and implications from a multiseason household transmission study},
	volume = {3},
	issn = {2752-6542},
	shorttitle = {Influenza virus shedding and symptoms},
	doi = {10.1093/pnasnexus/pgae338},
	language = {en},
	number = {9},
	urldate = {2026-08-24},
	journal = {PNAS Nexus},
	author = {Morris, Sinead E and Nguyen, Huong Q and Grijalva, Carlos G and Hanson, Kayla E and Zhu, Yuwei and Biddle, Jessica E and Meece, Jennifer K and Halasa, Natasha B and Chappell, James D and Mellis, Alexandra M and Reed, Carrie and Biggerstaff, Matthew and Belongia, Edward A and Talbot, H Keipp and Rolfes, Melissa A},
	editor = {Glymour, Medellena},
	month = sep,
	year = {2024},
	pages = {pgae338},
}

@article{ip_viral_2017,
	title = {Viral {Shedding} and {Transmission} {Potential} of {Asymptomatic} and {Paucisymptomatic} {Influenza} {Virus} {Infections} in the {Community}},
	volume = {64},
	issn = {1058-4838, 1537-6591},
	doi = {10.1093/cid/ciw841},
	language = {en},
	number = {6},
	urldate = {2026-08-24},
	journal = {Clinical Infectious Diseases},
	author = {Ip, Dennis K. M. and Lau, Lincoln L. H. and Leung, Nancy H. L. and Fang, Vicky J. and Chan, Kwok-Hung and Chu, Daniel K. W. and Leung, Gabriel M. and Peiris, J. S. Malik and Uyeki, Timothy M. and Cowling, Benjamin J.},
	month = mar,
	year = {2017},
	pages = {736--742},
}

@misc{Buchholz_GrippeWeb_Daten_2026,
author = {Buchholz, Udo and Lehfeld, Ann-Sophie and Loenenbach, Anna and Prahm, Kerstin and Preuß, Ute and Stepanovich-Falke, Anastassiya and Haas, Walter},
doi = {10.5281/zenodo.22024280},
month = aug,
title = {{GrippeWeb - Daten des Wochenberichts}},
url = {https://robert-koch-institut.github.io/GrippeWeb_Daten_des_Wochenberichts},
year = {2026}
}

@misc{a3m,
title = {{COVID-19 pandemic - Germany}},
author = {{A3M Global Monitoring GmbH}},
howpublished = {\url{https://global-monitoring.com/gm/page/events/epidemic-0001937.eP207vO5rIwe.html}},
year = {2025},
note = {\url{https://global-monitoring.com/gm/page/events/epidemic-0001937.eP207vO5rIwe.html}, accessed: 10 Aug 2026}
}

@article{jahnke_reduced_2011,
	title = {On {Reduced} {Models} for the {Chemical} {Master} {Equation}},
	volume = {9},
	issn = {1540-3459, 1540-3467},
	doi = {10.1137/110821500},
	language = {en},
	number = {4},
	urldate = {2026-08-25},
	journal = {Multiscale Modeling \& Simulation},
	author = {Jahnke, Tobias},
	month = oct,
	year = {2011},
	pages = {1646--1676},
}

@article{menz_hybrid_2012,
	title = {Hybrid {Stochastic}--{Deterministic} {Solution} of the {Chemical} {Master} {Equation}},
	volume = {10},
	issn = {1540-3459, 1540-3467},
	doi = {10.1137/110825716},
	language = {en},
	number = {4},
	urldate = {2026-08-25},
	journal = {Multiscale Modeling \& Simulation},
	author = {Menz, Stephan and Latorre, Juan C. and Schütte, Christof and Huisinga, Wilhelm},
	month = jan,
	year = {2012},
	pages = {1232--1262},
}

@article{schnoerr_comparison_2015,
	title = {Comparison of different moment-closure approximations for stochastic chemical kinetics},
	volume = {143},
	issn = {0021-9606, 1089-7690},
	doi = {10.1063/1.4934990},
	language = {en},
	number = {18},
	urldate = {2026-08-25},
	journal = {The Journal of Chemical Physics},
	author = {Schnoerr, David and Sanguinetti, Guido and Grima, Ramon},
	month = nov,
	year = {2015},
	pages = {185101},
}

@inproceedings{hespanha_moment_2008,
	address = {St. Julian's, Malta},
	title = {Moment closure for biochemical networks},
	isbn = {978-1-4244-1687-5},
	doi = {10.1109/ISCCSP.2008.4537208},
	urldate = {2026-08-25},
	booktitle = {2008 3rd {International} {Symposium} on {Communications}, {Control} and {Signal} {Processing}},
	publisher = {IEEE},
	author = {Hespanha, Joao},
	month = mar,
	year = {2008},
	pages = {142--147},
}

@article{bjornstad_seirs_2020,
	title = {The {SEIRS} model for infectious disease dynamics},
	volume = {17},
	copyright = {https://www.springer.com/tdm},
	issn = {1548-7091, 1548-7105},
	url = {https://www.nature.com/articles/s41592-020-0856-2},
	doi = {10.1038/s41592-020-0856-2},
	language = {en},
	number = {6},
	urldate = {2026-09-28},
	journal = {Nature Methods},
	author = {Bjørnstad, Ottar N. and Shea, Katriona and Krzywinski, Martin and Altman, Naomi},
	month = jun,
	year = {2020},
	pages = {557--558},
}

@article{andersonmodifiednext2007,
	title = {A modified next reaction method for simulating chemical systems with time dependent propensities and delays},
	volume = {127},
	issn = {0021-9606, 1089-7690},
	url = {https://pubs.aip.org/aip/jcp/article/349528},
	doi = {10.1063/1.2799998},
	language = {en},
	number = {21},
	urldate = {2023-05-15},
	journal = {The Journal of Chemical Physics},
	author = {Anderson, David F.},
	month = dec,
	year = {2007},
	pages = {214107},
}

@article{gillespie_moment-closure_2009,
	title = {Moment-closure approximations for mass-action models},
	volume = {3},
	doi = {10.1049/iet-syb:20070031},
	number = {1},
	urldate = {2026-08-10},
	journal = {IET Systems Biology},
	publisher = {The Institution of Engineering and Technology},
	author = {Gillespie, C.S.},
	month = jan,
	year = {2009},
	pages = {52--58},
}

@article{mcquarrie1967stochastic,
  title={Stochastic approach to chemical kinetics},
  author={McQuarrie, Donald A},
  journal={Journal of applied probability},
  volume={4},
  number={3},
  pages={413--478},
  year={1967},
  publisher={Cambridge University Press}
}

@article{kehrer_hybrid_2027,
	title = {A hybrid {ABM}-{PDE} framework for real-world infectious disease simulations},
	volume = {162},
	issn = {0307-904X},
	doi = {10.1016/j.apm.2026.117173},
	urldate = {2026-08-11},
	journal = {Applied Mathematical Modelling},
	author = {Kehrer, Kristina and Conrad, Tim O. F.},
	month = feb,
	year = {2027},
	pages = {117173},
}

@article{wendler_high_order,
	title   = {Higher order schemes for the numerical solution of integro-differential equation-based age-of-infection models with time-dependent contact rates},
	author  = {Wendler, Anna and Hasenauer, Jan and Kühn, Martin},
	year    = {2026},
	note     = { Submitted.}
}

@article{ZUNKER2027_stage_aligned,
title = {Efficient numerical computation of traveler states in explicit mobility-based metapopulation models: Mathematical theory and application to epidemics},
journal = {Mathematics and Computers in Simulation},
volume = {251},
pages = {798-812},
year = {2027},
issn = {0378-4754},
doi = {10.1016/j.matcom.2026.08.029},
author = {Henrik Zunker and René Schmieding and Jan Hasenauer and Martin J. Kühn},
}

@article{kwak2026hybridepidemicsimulationframework,
      title={Hybrid epidemic simulation framework coupling equation-based and individual-based models}, 
      author={Jaeyoung Kwak and Michael H. Lees and Chin Chun Ooi and Wentong Cai},
      year={2026},
      doi={10.48550/arXiv.2609.35162},
      journal={arxiv}
}
	
	\appendix
	\newgeometry{top=1.5cm, bottom=1.5cm, left=1.5cm, right=1.5cm}
	\newpage
	\begin{center}
		\Large{Supplementary Information}
	\end{center}
	
	\section{Moment equations}
	\label{sec:supp_moment_eq}
	For $l=1,\ldots, n_R$, the evolution equations for the mean of the SIRS-type metapopulation model are given by
	\begin{align*}
		\frac{d\mu_{\iota(S,l)}}{dt} &= -\lambda^{(l)}\cdot \biggl(\mu_{\iota(S,l)}\mu_{\iota(I,l)}+\mathbb{E}\left[(X_S^{(l)}-\mu_{\iota(S,l)})(X_I^{(l)}-\mu_{\iota(I,l)})\right]\biggr) + \alpha \mu_{\iota(R,l)} + \sum_{\substack{k=1 \\ k \neq l}}^{n_R}( \kappa_S^{(k,l)}\mu_{\iota(S,k)}-\kappa_S^{(l,k)}\mu_{\iota(S,l)})\\
		\frac{d\mu_{\iota(I,l)}}{dt}&=\lambda^{(l)}\cdot\biggl(\mu_{\iota(S,l)}\mu_{\iota(I,l)}+\mathbb{E}\left[(X_S^{(l)}-\mu_{\iota(S,l)})(X_I^{(l)}-\mu_{\iota(I,l)})\right]\biggr) -\gamma \mu_{\iota(I,l)} + \sum_{\substack{k=1 \\ k \neq l}}^{n_R}(\kappa_I^{(k,l)}\mu_{\iota(I,k)}-\kappa_I^{(l,k)}\mu_{\iota(I,l)})\\
		\frac{d\mu_{\iota(R,l)}}{dt}&=\gamma \mu_{\iota(I,l)} - \alpha \mu_{\iota(R,l)} + \sum_{\substack{k=1 \\ k \neq l}}^{n_R}(\kappa_R^{(k,l)}\mu_{\iota(R,k)}-\kappa_R^{(l,k)}\mu_{\iota(R,l)})
	\end{align*}
	The ODE system for the variances $\sigma_{\iota(z,l),\iota(z,l)}=\mathbb{E}[(X_z^{(l)}-\mu_{\iota(z,l)})^2]$, $z\in\{S,I,R\}$, is given by
	\begin{align*}
		\frac{d\sigma_{\iota(S,l),\iota(S,l)}}{dt}&=\lambda^{(l)}\mu_{\iota(S,l)}\mu_{\iota(I,l)}+\alpha\mu_{\iota(R,l)}+\sum_{\substack{k=1 \\ k \neq l}}^{n_R}(\kappa_S^{(l,k)}\mu_{\iota(S,l)}+\kappa_S^{(k,l)}\mu_{\iota(S,k)})\\
		&-2\lambda^{(l)}\biggl(\mu_{\iota(I,l)}\sigma_{\iota(S,l),\iota(S,l)}+\mu_{\iota(S,l)}\mathbb{E}[(X_S^{(l)}-\mu_{\iota(S,l)})(X_I^{(l)}-\mu_{\iota(I,l)})]\biggr)\\
		&+2\alpha\mathbb{E}[(X_S^{(l)}-\mu_{\iota(S,l)})(X_R^{(l)}-\mu_{\iota(R,l)})]
		\\&+2\sum_{\substack{k=1 \\ k \neq l}}^{n_R} \left(-\kappa_S^{(l,k)}\sigma_{\iota(S,l),\iota(S,l)}+\kappa_S^{(k,l)}\mathbb{E}[(X_S^{(l)}-\mu_{\iota(S,l)})(X_S^{(k)}-\mu_{\iota(S,k)})]\right)\\
		&+\lambda^{(l)}\mathbb{E}[(X_S^{(l)}-\mu_{\iota(S,l)})(X_I^{(l)}-\mu_{\iota(I,l)})]-2\lambda^{(l)}\mathbb{E}[(X_S^{(l)}-\mu_{\iota(S,l)})^2(X_I^{(l)}-\mu_{\iota(I,l)})]\\
		\frac{d\sigma_{\iota(I,l),\iota(I,l)}}{dt}&=\lambda^{(l)}\mu_{\iota(S,l)}\mu_{\iota(I,l)}+\gamma \mu_{I^{(l)}}+\sum_{\substack{k=1 \\ k \neq l}}^{n_R}(\kappa_I^{(l,k)}\mu_{\iota(I,l)}+\kappa_I^{(k,l)}\mu_{\iota(I,k)})\\
		&+2\lambda^{(l)}\biggl(\mu_{\iota(I,l)}\mathbb{E}[(X_S^{(l)}-\mu_{\iota(S,l)})(I^{(l)}-\mu_{\iota(I,l)})]+\mu_{\iota(S,l)}\sigma_{\iota(I,l),\iota(I,l)}\biggr)-2\gamma\sigma_{\iota(I,l),\iota(I,l)}\\
		&+2\sum_{\substack{k=1 \\ k \neq l}}^{n_R} \left(-\kappa_I^{(l,k)}\sigma_{\iota(I,l),\iota(I,l)}+\kappa_I^{(k,l)}\mathbb{E}[(X_I^{(l)}-\mu_{\iota(I,l)})(X_I^{(k)}-\mu_{\iota(I,k)})]\right)\\
		&+\lambda^{(l)}\mathbb{E}[(X_S^{(l)}-\mu_{\iota(S,l)})(X_I^{(l)}-\mu_{\iota(I,l)})]+2\lambda^{(l)}\mathbb{E}[(X_S^{(l)}-\mu_{\iota(S,l)})(X_I^{(l)}-\mu_{\iota(I,l)})^2]\\
		\frac{d\sigma_{\iota(R,l),\iota(R,l)}}{dt}&=\gamma\mu_{\iota(I,l)}+\alpha\mu_{\iota(R,l)}+\sum_{\substack{k=1 \\ k \neq l}}^{n_R}(\kappa_R^{(l,k)}\mu_{\iota(R,l)}+\kappa_R^{(k,l)}\mu_{\iota(R,k)})+2\gamma\mathbb{E}[(X_I^{(l)}-\mu_{\iota(I,l)})(X_R^{(l)}-\mu_{\iota(R,l)})]\\ &-2\alpha\sigma_{\iota(R,l),\iota(R,l)}+2\sum_{\substack{k=1 \\ k \neq l}}^{n_R} \left(-\kappa_R^{(l,k)}\sigma_{\iota(R,l),\iota(R,l)}+\kappa_R^{(k,l)}\mathbb{E}[(X_R^{(l)}-\mu_{\iota(R,l)})(X_R^{(k)}-\mu_{\iota(R,k)})]\right)
	\end{align*}
	In general, we denote the $m$-th central moment of the SIRS-type metapopulation model as
	\begin{align*}
		M_{i_S^{(1)},i_I^{(1)}, i_R^{(1)}, \ldots, i_R^{(n_R)}}=\mathbb{E}[(X_S^{(1)}-\mu_{\iota(S,1)})^{i_S^{(1)}}(X_I^{(1)}-\mu_{\iota(I,1)})^{i_I^{(1)}}(X_R^{(1)}-\mu_{\iota(R,1)})^{i_R^{(1)}}\ldots(X_R^{(n_R)}-\mu_{\iota(R,n_R)})^{i_R^{(n_R)}}]
	\end{align*}
	where $m=i_S^{(1)}+i_I^{(1)}+\ldots+i_R^{(n_R)}$. 
	Moments of order $m<2$ are given by
	\begin{align*}
		M_0&=\mathbb{E}[1]=1\\
		M_1&=\mathbb{E}[(X_z^{(l)}-\mu_{\iota(z,l)})]=0, \ \forall z\in\{S,I,R\}
	\end{align*}
	Based on~\cite{lee_moment_2009}, $M_{i_S^{(1)},i_I^{(1)}, i_R^{(1)}, \ldots, i_R^{(n_R)}}$ of order $m\geq 2$ is given by
	\begin{align}
		\label{eq:general_moment_eq}
		\begin{split}
			&\frac{dM_{i_S^{(1)},i_I^{(1)}, i_R^{(1)}, \ldots, i_R^{(n_R)}}}{dt}\\ &=\sum_{l=1}^{n_R}\biggl[\lambda^{(l)}\mu_{\iota(S,l)}\mu_{\iota(I,l)}\sum_{h_S^{(l)}+h_I^{(l)} \neq i_S^{(l)}+i_I^{(l)}} \binom{i_S^{(l)}}{h_S^{(l)}}\binom{i_I^{(l)}}{h_I^{(l)}} (-1)^{i_S^{(l)}-h_S^{(l)}} M_{i_S^{(1)},\ldots, h_S^{(l)},h_I^{(l)},\ldots, i_R^{(n_R)}}\\
			&+ \gamma \mu_{\iota(I,l)} \sum_{h_I^{(l)}+h_R^{(l)} \neq i_I^{(l)}+i_R^{(l)}} \binom{i_I^{(l)}}{h_I^{(l)}}\binom{i_R^{(l)}}{h_R^{(l)}} (-1)^{i_I^{(l)}-h_I^{(l)}} M_{i_S^{(1)},\ldots, h_I^{(l)},h_R^{(l)},\ldots, i_R^{(n_R)}}\\
			&+ \alpha \mu_{\iota(R,l)} \sum_{h_S^{(l)}+h_R^{(l)} \neq i_S^{(l)}+i_R^{(l)}} \binom{i_S^{(l)}}{h_S^{(l)}}\binom{i_R^{(l)}}{h_R^{(l)}} (-1)^{i_R^{(l)}-h_R^{(l)}} M_{i_S^{(1)},\ldots, h_S^{(l)},i_I^{(l)},h_R^{(l)},\ldots, i_R^{(n_R)}}\\
			&+\sum_{k=l+1}^{n_R}\biggl(\sum_{h_S^{(l)}+h_S^{(k)}\neq i_S^{(l)}+i_S^{(k)}}\binom{i_S^{(l)}}{h_S^{(l)}}\binom{i_S^{(k)}}{h_S^{(k)}}M_{i_S^{(1)},\ldots, h_S^{(l)},\ldots, h_S^{(k)},\ldots, i_R^{(n_R)}} \Big((-1)^{i_S^{(l)}-h_S^{(l)}}\kappa_S^{(l,k)}\mu_{\iota(S,l)} + (-1)^{i_S^{(k)}-h_S^{(k)}}\kappa_S^{(k,l)}\mu_{\iota(S,k)}\Big) \\
			&+ \sum_{h_I^{(l)}+h_I^{(k)}\neq i_I^{(l)}+i_I^{(k)}}\binom{i_I^{(l)}}{h_I^{(l)}}\binom{i_I^{(k)}}{h_I^{(k)}}M_{i_S^{(1)},\ldots, h_I^{(l)},\ldots, h_I^{(k)},\ldots, i_R^{(n_R)}}\Big((-1)^{i_I^{(l)}-h_I^{(l)}}\kappa_I^{(l,k)}\mu_{\iota(I,l)}+ (-1)^{i_I^{(k)}-h_I^{(k)}}\kappa_I^{(k,l)}\mu_{\iota(I,k)}\Big)\\
			&+\sum_{h_R^{(l)}+h_R^{(k)}\neq i_R^{(l)}+i_R^{(k)}}\binom{i_R^{(l)}}{h_R^{(l)}}\binom{i_R^{(k)}}{h_R^{(k)}}M_{i_S^{(1)},\ldots, h_R^{(l)},\ldots, h_R^{(k)},\ldots, i_R^{(n_R)}} \Big((-1)^{i_R^{(l)}-h_R^{(l)}}\kappa_R^{(l,k)}\mu_{\iota(R,l)} + (-1)^{i_R^{(k)}-h_R^{(k)}}\kappa_R^{(k,l)}\mu_{\iota(R,k)}\Big)\biggr)\\
			&+\sum_{h_S^{(l)}+h_I^{(l)}\neq i_S^{(l)}+i_I^{(l)}}\binom{i_S^{(l)}}{h_S^{(l)}}\binom{i_I^{(l)}}{h_I^{(l)}} (-1)^{i_S^{(l)}-h_S^{(l)}}\cdot\biggl(\Big( \lambda^{(l)}\mu_{\iota(I,l)} M_{i_S^{(1)},\ldots, h_S^{(l)}+1,h_I^{(l)},\ldots, i_R^{(n_R)}}\\
			&+ \lambda^{(l)}\mu_{\iota(S,l)} M_{i_S^{(1)},\ldots, h_S^{(l)},h_I^{(l)}+1,\ldots, i_R^{(n_R)}}\Big)\biggr) +\gamma \sum_{h_I^{(l)}+h_R^{(l)} \neq i_I^{(l)}+i_R^{(l)}} \binom{i_I^{(l)}}{h_I^{(l)}}\binom{i_R^{(l)}}{h_R^{(l)}} (-1)^{i_I^{(l)}-h_I^{(l)}} M_{i_S^{(1)},\ldots, h_I^{(l)}+1,h_R^{(l)},\ldots, i_R^{(n_R)}}\\ 
			&+\alpha \sum_{h_S^{(l)}+h_R^{(l)} \neq i_S^{(l)}+i_R^{(l)}} \binom{i_S^{(l)}}{h_S^{(l)}}\binom{i_R^{(l)}}{h_R^{(l)}} (-1)^{i_R^{(l)}-h_R^{(l)}} M_{i_S^{(1)},\ldots, h_S^{(l)},i_I^{(l)},h_R^{(l)}+1,\ldots, i_R^{(n_R)}}\\
			&+\sum_{k=l+1}^{n_R}\biggl(\sum_{h_S^{(l)}+h_S^{(k)}\neq i_S^{(l)}+i_S^{(k)}}\binom{i_S^{(l)}}{h_S^{(l)}}\binom{i_S^{(k)}}{h_S^{(k)}}\Big((-1)^{i_S^{(l)}-h_S^{(l)}}\kappa_S^{(l,k)}M_{i_S^{(1)},\ldots, h_S^{(l)}+1,\ldots, h_S^{(k)},\ldots, i_R^{(n_R)}} \\
			&+ (-1)^{i_S^{(k)}-h_S^{(k)}}\kappa_S^{(k,l)}M_{i_S^{(1)},\ldots, h_S^{(l)},\ldots, h_S^{(k)}+1,\ldots, i_R^{(n_R)}}\Big)\\
			&+\sum_{h_I^{(l)}+h_I^{(k)}\neq i_I^{(l)}+i_I^{(k)}}\binom{i_I^{(l)}}{h_I^{(l)}}\binom{i_I^{(k)}}{h_I^{(k)}}
			\Big((-1)^{i_I^{(l)}-h_I^{(l)}}\kappa_I^{(l,k)}M_{i_S^{(1)},\ldots, h_I^{(l)}+1,\ldots, h_I^{(k)},\ldots, i_R^{(n_R)}}\\
			&+ (-1)^{i_I^{(k)}-h_I^{(k)}}\kappa_I^{(k,l)}M_{i_S^{(1)},\ldots, h_I^{(l)},\ldots, h_I^{(k)}+1,\ldots, i_R^{(n_R)}}\Big)
			+\sum_{h_R^{(l)}+h_R^{(k)}\neq i_R^{(l)}+i_R^{(k)}}\binom{i_R^{(l)}}{h_R^{(l)}}\binom{i_R^{(k)}}{h_R^{(k)}}\\
			&\Big((-1)^{i_R^{(l)}-h_R^{(l)}}\kappa_R^{(l,k)}M_{i_S^{(1)},\ldots, h_R^{(l)}+1,\ldots, h_R^{(k)},\ldots, i_R^{(n_R)}}+ (-1)^{i_R^{(k)}-h_R^{(k)}}\kappa_R^{(k,l)}M_{i_S^{(1)},\ldots, h_R^{(l)},\ldots, h_R^{(k)}+1,\ldots, i_R^{(n_R)}}\Big)\biggr)\\
			&+\lambda^{(l)}\sum_{h_S^{(l)}+h_I^{(l)} \neq i_S^{(l)}+i_I^{(l)}} \binom{i_S^{(l)}}{h_S^{(l)}}\binom{i_I^{(l)}}{h_I^{(l)}} (-1)^{i_S^{(l)}-h_S^{(l)}}M_{i_S^{(1)},\ldots, h_S^{(l)}+1,h_I^{(l)}+1,\ldots, i_R^{(n_R)}}\\
			&-i_S^{(l)}\mu_{\iota(S,l)}'M_{i_S^{(1)},\ldots, i_S^{(l)}-1,\ldots, i_R^{(n_R)}}-i_I^{(l)}\mu_{\iota(I,l)}'M_{i_S^{(1)},\ldots, i_I^{(l)}-1,\ldots, i_R^{(n_R)}}-i_R^{(l)}\mu_{\iota(R,l)}'M_{i_S^{(1)},\ldots, i_R^{(l)}-1,\ldots, i_R^{(n_R)}}\biggr]
		\end{split}
	\end{align}
	
	\newpage
	\section{Supplementary Algorithms}
	
	\begin{algorithm}[!ht]
		\caption{Temporal-spatial hybrid simulation}
		\label{alg:temporal_spatial_hybrid} 
		\footnotesize
		Initialize stochastic model for all regions, moment equations with zeros, $\Delta t$ with $\Delta t_{sw}=\Delta t_{ex}$ and $t$ with $t_0$\\
		\textbf{While} $t<t_\text{max}$\\
		\hspace{0.5cm}\\
		\hspace{0.5cm}\textbf{Foreach} stochastically modeled region $k$\\
		\hspace{1cm}\textbf{If} $\Gamma^{(S\rightarrow D)}\left(\mathbf{X}\right)=1$\\
		\hspace{1.5cm}\textbf{Forall} infection states $z\in\{S,I,R\}$\\
		\hspace{2cm}$\mu^{D}_{\iota(z,k)}(t)=\mu^{S}_{\iota(z,k)}(t)$\\
		\hspace{2cm}\textbf{Forall} infection states $y\in\{S,I,R\}$\\
		\hspace{2.5cm}$\sigma^{D}_{\iota(y,k),\iota(z,k)}(t)=\sigma^{S}_{\iota(y,k),\iota(z,k)}(t)$\\
		\hspace{2cm}\textbf{Forall} $s=1,\ldots,n_{sims}$\\
		\hspace{2.5cm}$X_{z,s}^{(k)}=0$\\
		\hspace{1.5cm}Set infection state transition and outgoing spatial transition rates\\
		\hspace{1.5cm}for region $k$ in moment equation-based model\\
		\hspace{1.5cm}Remove infection state transition and outgoing spatial transition rates\\
		\hspace{1.5cm}for region $k$ in stochastic model\\
		
		\begin{tikzpicture}[overlay,remember picture]
			\draw[decorate,decoration={brace,amplitude=8pt}, line width=1.5pt, draw=olive!50!green!100!]
			([xshift=12cm,yshift=4cm]pic cs:cbeg) --
			([xshift=12cm,yshift=0.2cm]pic cs:cend)
			node[midway,right=8pt, align=left]{\color{olive!50!green!100!}Convert stochastic\\\color{olive!50!green!100!}to MoM};
		\end{tikzpicture}
		
		\hspace{0.5cm}\textbf{Foreach} region $l$ modeled with moment equations\\
		\hspace{1cm}\textbf{If} $\Gamma^{(D\rightarrow S)}\left(\left[\mu^D(t),\Sigma^D(t)\right]^\top\right)=1$\\
		\hspace{1.5cm}\textbf{Forall} infection states $z\in\{S,I,R\}$\\
		\hspace{2cm}$\xi_{\iota(z,l)}=0$\\
		\hspace{2.5cm}\textbf{Forall} $s=1,...,n_{sims}$\\
		\hspace{3cm}$\tilde{X}^{(l)}_{z,s} \sim \mathcal{N}(\mu^{D}_{\iota(z,l)}, {\sigma_{\iota(z,l),\iota(z,l)}^{D}})$\\
		\hspace{3cm}$X^{(l)}_{z,s}=\max\{0,\lfloor\tilde{X}^{(l)}_{z,s}+\xi_{\iota(z,l)}\rceil\}$\\
		\hspace{3cm}$\xi_{\iota(z,l)}=\tilde{X}^{(l)}_{z,s}-X^{(l)}_{z,s}$\\
		\hspace{2.5cm}$\mu^{D}_{\iota(z,l)}(t)=0$\\
		\hspace{2cm}\textbf{Forall} infection states $y\in\{S,I,R\}$\\
		\hspace{2.5cm}$\sigma^{D}_{\iota(y,l),\iota(z,l)}(t)=0$\\
		\hspace{1.5cm}Set infection state transition and outgoing spatial transition rates\\
		\hspace{1.5cm}for region $l$ in stochastic model\\
		\hspace{1.5cm}Remove infection state transition and outgoing spatial transition rates\\
		\hspace{1.5cm}for region $l$ in moment equation-based model\\
		
		\begin{tikzpicture}[overlay,remember picture]
			\draw[decorate,decoration={brace,amplitude=8pt}, line width=1.5pt, draw=olive!50!green!100!]
			([xshift=12cm,yshift=5.5cm]pic cs:cbeg) --
			([xshift=12cm,yshift=0.2cm]pic cs:cend)
			node[midway,right=8pt, align=left]{\color{olive!50!green!100!}Convert MoM\\\color{olive!50!green!100!}to stochastic model};
		\end{tikzpicture}
		
		\hspace{0.5cm}$t=\min\{t_{max},t+\Delta t\}$\\
		\hspace{0.5cm}Advance both models until $t$\\
		\hspace{0.5cm}\\
		\hspace{0.5cm}\textbf{Foreach} stochastically modeled region $k$\\
		\hspace{1cm}\textbf{Forall} infection states $z\in\{S,I,R\}$\\
		\hspace{1.5cm}\textbf{If} $\mu^{D}_{\iota(z,k)}(t)>0$\\
		\hspace{2cm}$\xi_{\iota(z,k)}=0$\\
		\hspace{2cm}\textbf{Forall} $s=1,\ldots,n_{sims}$\\
		\hspace{2.5cm}$\tilde{X}^{(k)}_{z,s} \sim \mathcal{N}(\mu^{D}_{\iota(z,k)}, {\sigma_{\iota(z,k),\iota(z,k)}^{D}})$\\
		\hspace{2.5cm}$X^{(k)}_{z,s}+=\max\{0,\lfloor\tilde{X}^{(k)}_{z,s}+\xi_{\iota(z,k)}\rceil\}$\\
		\hspace{2.5cm}$\xi_{\iota(z,k)}=\tilde{X}^{(k)}_{z,s}-X^{(k)}_{z,s}$\\
		\hspace{2cm}$\mu^{D}_{\iota(z,k)}(t)=0$\\
		\hspace{2cm}\textbf{Forall} infection states $y\in\{S,I,R\}$\\
		\hspace{2.5cm}$\sigma^{D}_{\iota(y,k),\iota(z,k)}(t)=0$\\
		
		\begin{tikzpicture}[overlay,remember picture]
			\draw[decorate,decoration={brace,amplitude=8pt}, line width=1.5pt, draw=olive!50!green!100!]
			([xshift=12cm,yshift=4cm]pic cs:cbeg) --
			([xshift=12cm,yshift=0.2cm]pic cs:cend)
			node[midway,right=8pt, align=left]{\color{olive!50!green!100!}Exchange subpopulations\\\color{olive!50!green!100!} from moment equations\\\color{olive!50!green!100!}to stochastic model};
		\end{tikzpicture}
		
		\hspace{0.5cm}\textbf{Foreach} region $l$ modeled with moment equations\\
		\hspace{1cm}\textbf{Foreach} infection state $z\in\{S,I,R\}$\\
		\hspace{1.5cm}\textbf{If} $\mu^{S}_{\iota(z,l)}(t)>0$\\
		\hspace{2cm}$\mu^{D}_{\iota(z,l)}(t)+=\mu^{S}_{\iota(z,l)}(t)$\\
		\hspace{2cm}${\sigma_{\iota(z,l),\iota(z,l)}^{D}} \sim U\biggl(\left(\sqrt{\sigma_{\iota(z,l),\iota(z,l)}^{D}}-\sqrt{\sigma_{\iota(z,l),\iota(z,l)}^{S}}\right)^2, \left(\sqrt{\sigma_{\iota(z,l),\iota(z,l)}^{D}}+\sqrt{\sigma_{\iota(z,l),\iota(z,l)}^{S}}\right)^2\biggr)$\\
		\hspace{2cm}\textbf{Forall} $s=1,...,n_{sims}$\\
		\hspace{2.5cm}$X^{(l)}_{z,s}=0$\\
		\begin{tikzpicture}[overlay,remember picture]
			\draw[decorate,decoration={brace,amplitude=8pt}, line width=1.5pt, draw=olive!50!green!100!]
			([xshift=13.5cm,yshift=2.5cm]pic cs:cbeg) --
			([xshift=13.5cm,yshift=0.2cm]pic cs:cend)
			node[midway,right=8pt, align=left]{\color{olive!50!green!100!}Exchange subpopulations\\\color{olive!50!green!100!} from stochastic model\\\color{olive!50!green!100!}to moment equations};
		\end{tikzpicture}   
	\end{algorithm}
	
	\newpage
	\FloatBarrier
	
	\section{Supplementary Figures}
	
	\begin{figure}[!ht]
		\centering
		\includegraphics[width=\linewidth]{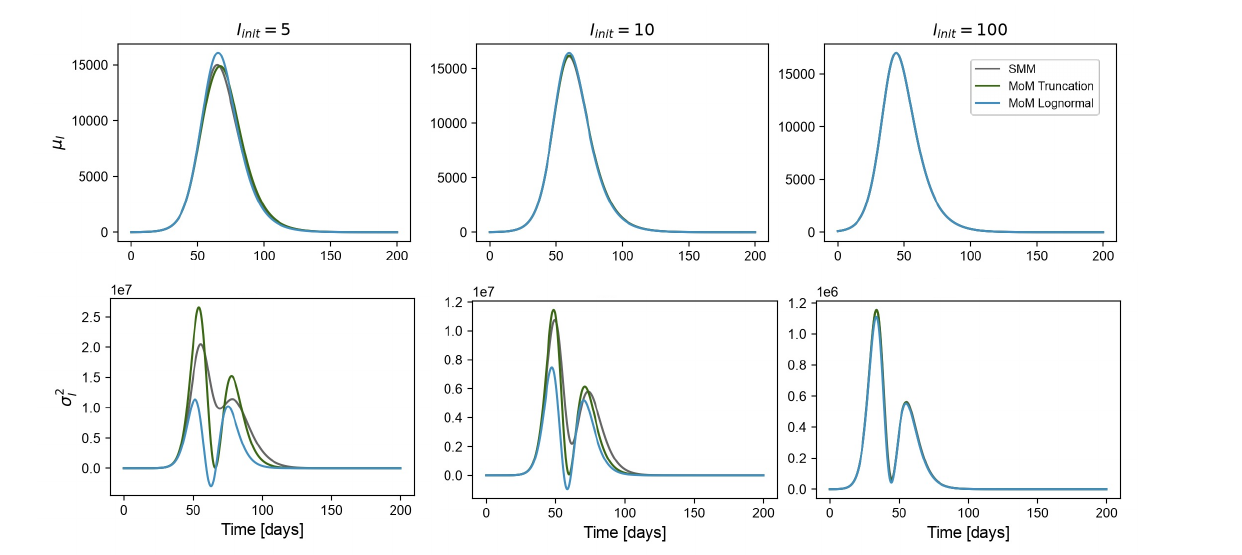}
		\caption{\textbf{Comparison of truncation closure and lognormal distributional closure.} Both closure methods approximate the mean (top) and the variance (bottom) differently well for varying initial conditions. The ground truth values (SMM) are obtained from $10,000$ Gillespie simulations of the stochastic model.}
		\label{fig:supp_moment_closures}
	\end{figure}
	
	\begin{figure}[!ht]
		\centering
		\includegraphics[width=0.92\linewidth]{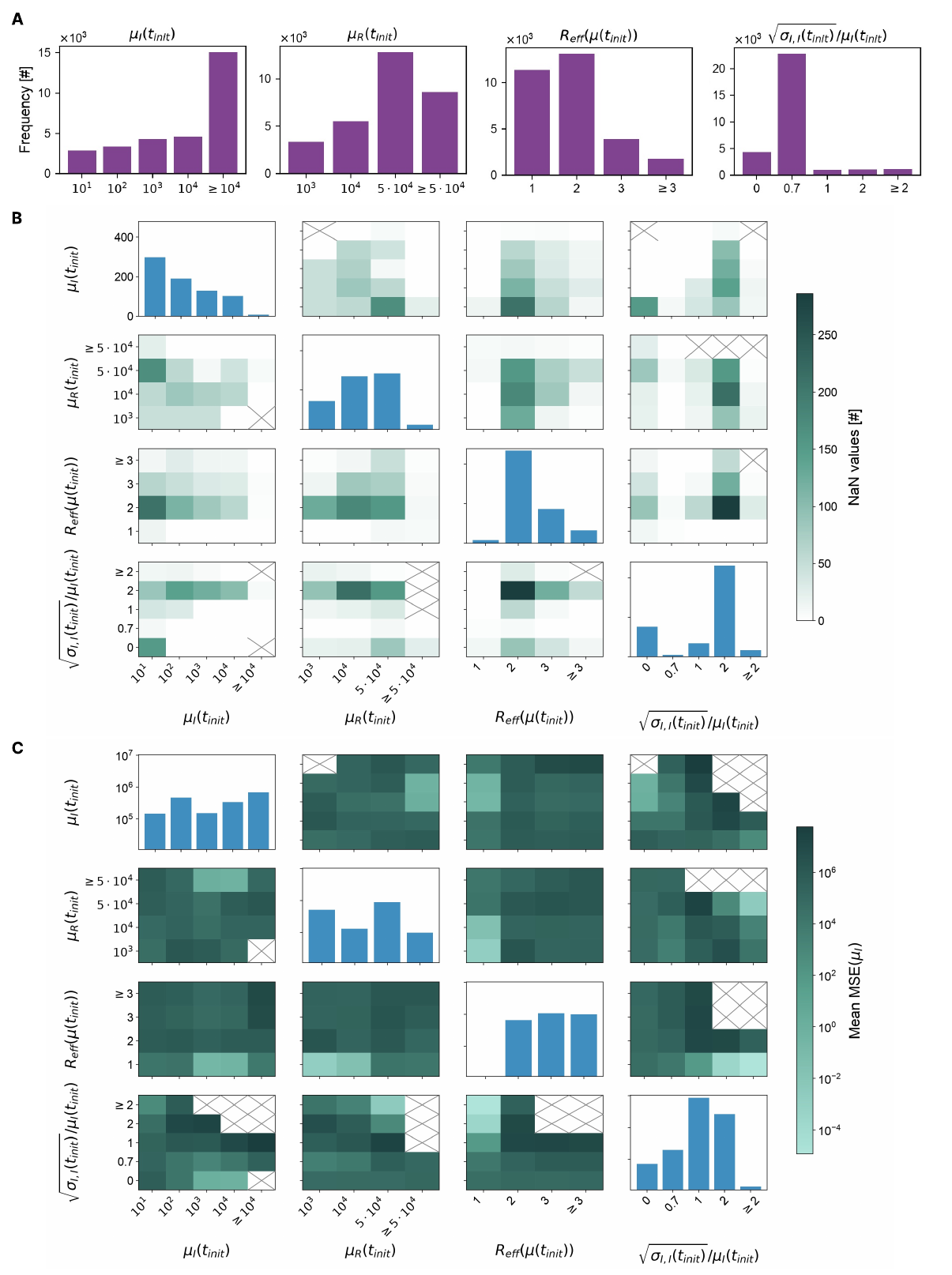}
		\caption{\textbf{Performance analysis of the MoM for the SIRS model.} For the analysis $50,000$ stochastic runs were compared to the solution of the moment equations using truncation closure for each parameter set. \textbf{A:} Distribution of initial conditions. \textbf{B:} Distribution of NaN value counts. \textbf{C:} Mean MSE of the simulated mean number of infected $\mu_I$. }
		\label{fig:supp_parameter_study_SIRS}
	\end{figure}
	
	\begin{figure}[!ht]
		\centering
		\includegraphics[width=\linewidth]{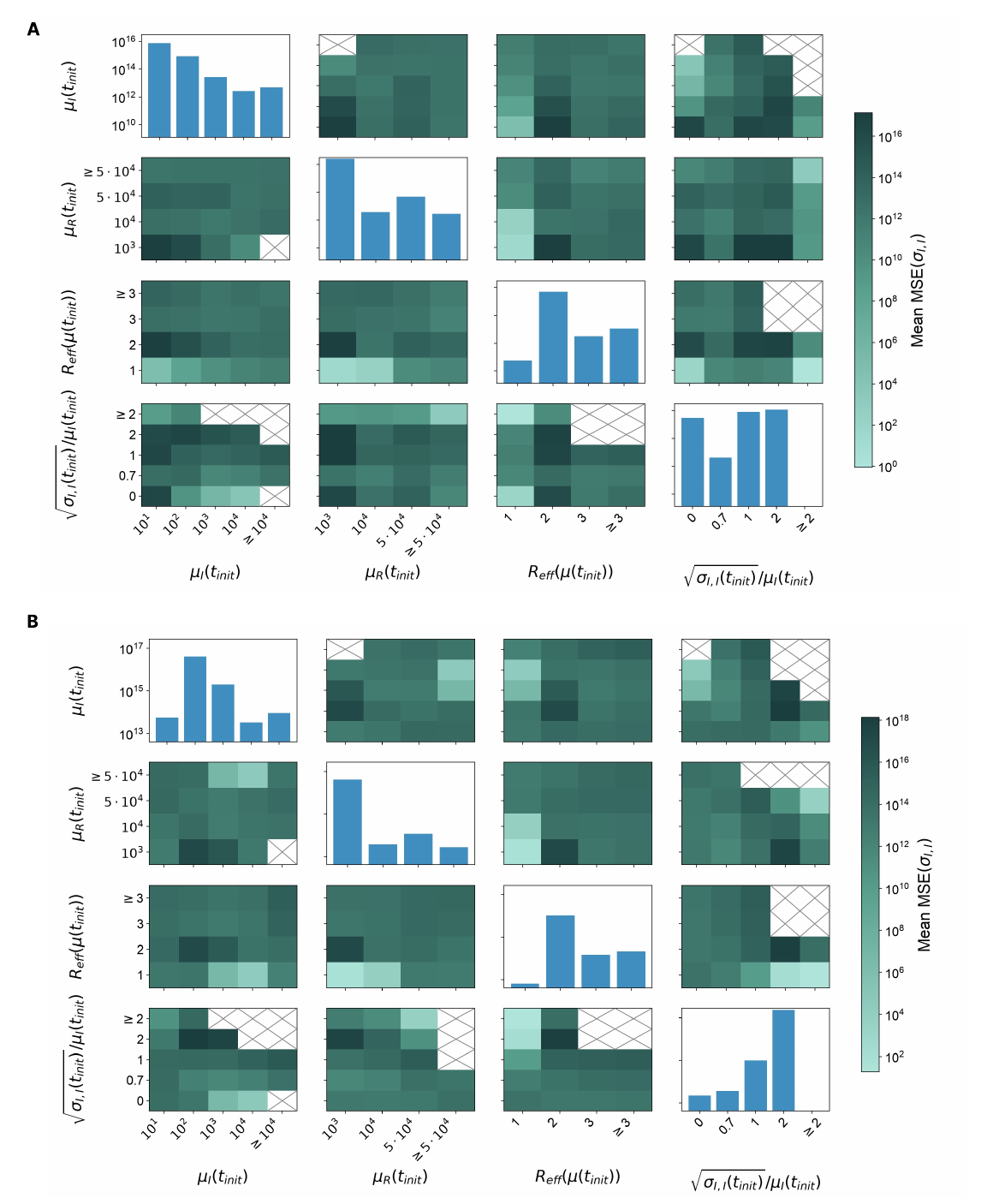}
		\caption{\textbf{Performance analysis of the MoM for the variance of the SIR and SIRS model.} Mean MSE of the simulated variance of the infected compartment $\sigma_I$ for \textbf{A:} the SIR model and \textbf{B:} the SIRS model.}
		\label{fig:supp_parameter_study_var}
	\end{figure}
	
	\begin{figure}[!ht]
		\centering
		\includegraphics[width=\linewidth]{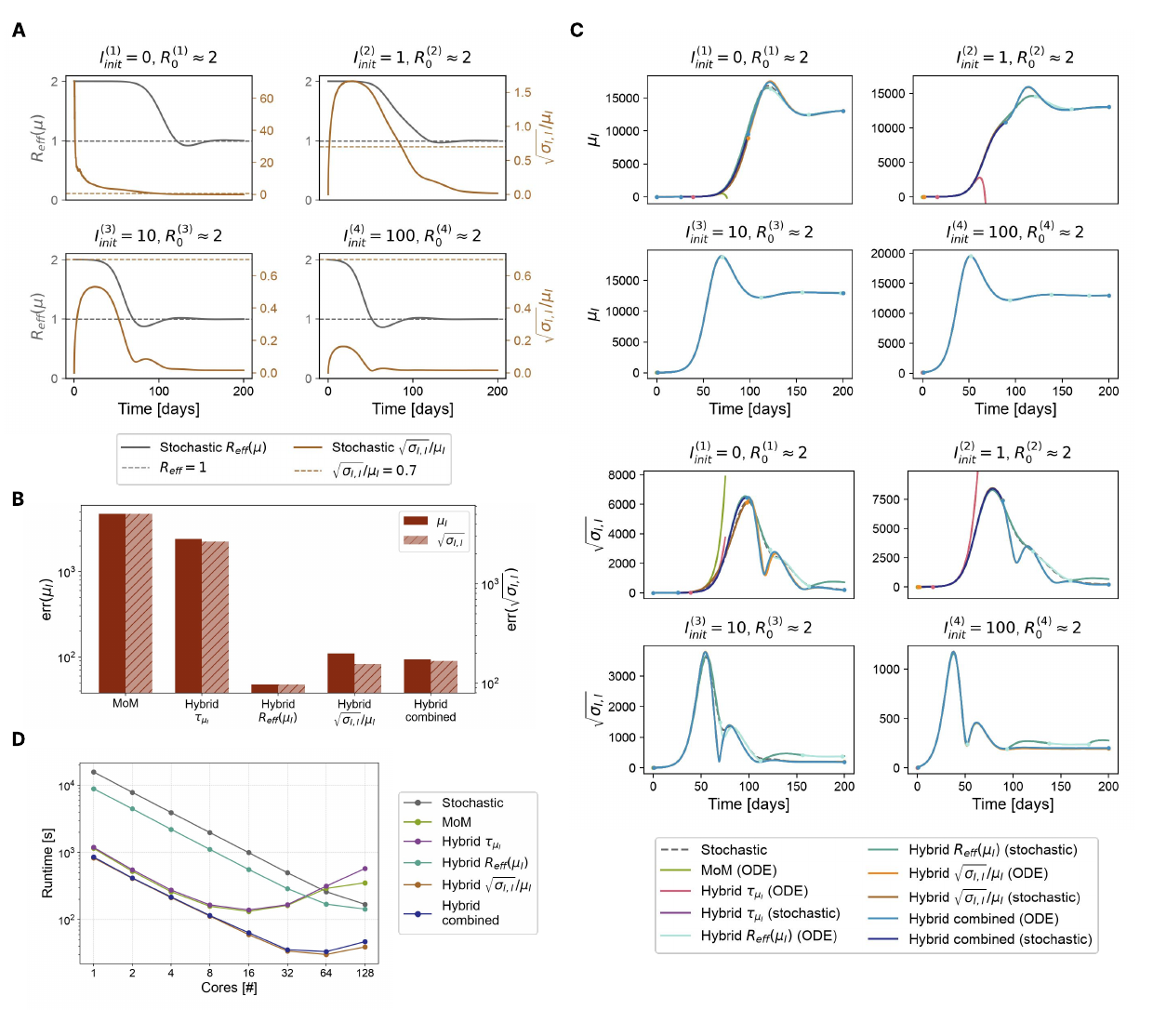}
		\caption{\textbf{Results of different condition functions for the SIRS-type model with four regions and a varying number of initially infected per region.} Shown are the results of $10,000$ simulations for each model. \textbf{A:} Temporal evolution of the relative standard deviation of the $I$ compartment and the reproduction number of the mean for each region. The horizontal red lines indicate the threshold values used for the corresponding conditions. \textbf{B:} Mean absolute error in mean and standard deviation of the deterministic model and the hybrid model with four different conditions. \textbf{C:} Evolution of mean (top) and standard deviation (bottom) in every region for the stochastic model, the deterministic model and the hybrid model with four different conditions. For the hybrid model the stochastic parts are displayed in a dark and the deterministic parts in a light color shade. \textbf{D:} Strong scaling of the stochastic model (Stochastic), the moment-closure approximation (MoM), and the hybrid models using the absolute threshold condition ($\tau_{\mu_I}$), the reproduction number condition ($R_{eff}(\mu_I)$), the standard deviation relation condition ($\frac{\sqrt{\sigma_{I,I}}}{\mu_I}$) and the combined condition.}
		\label{fig:hybrid_results_SIRS_I0}
	\end{figure}
	
	\begin{figure}[!ht]
		\centering
		\includegraphics[width=\linewidth]{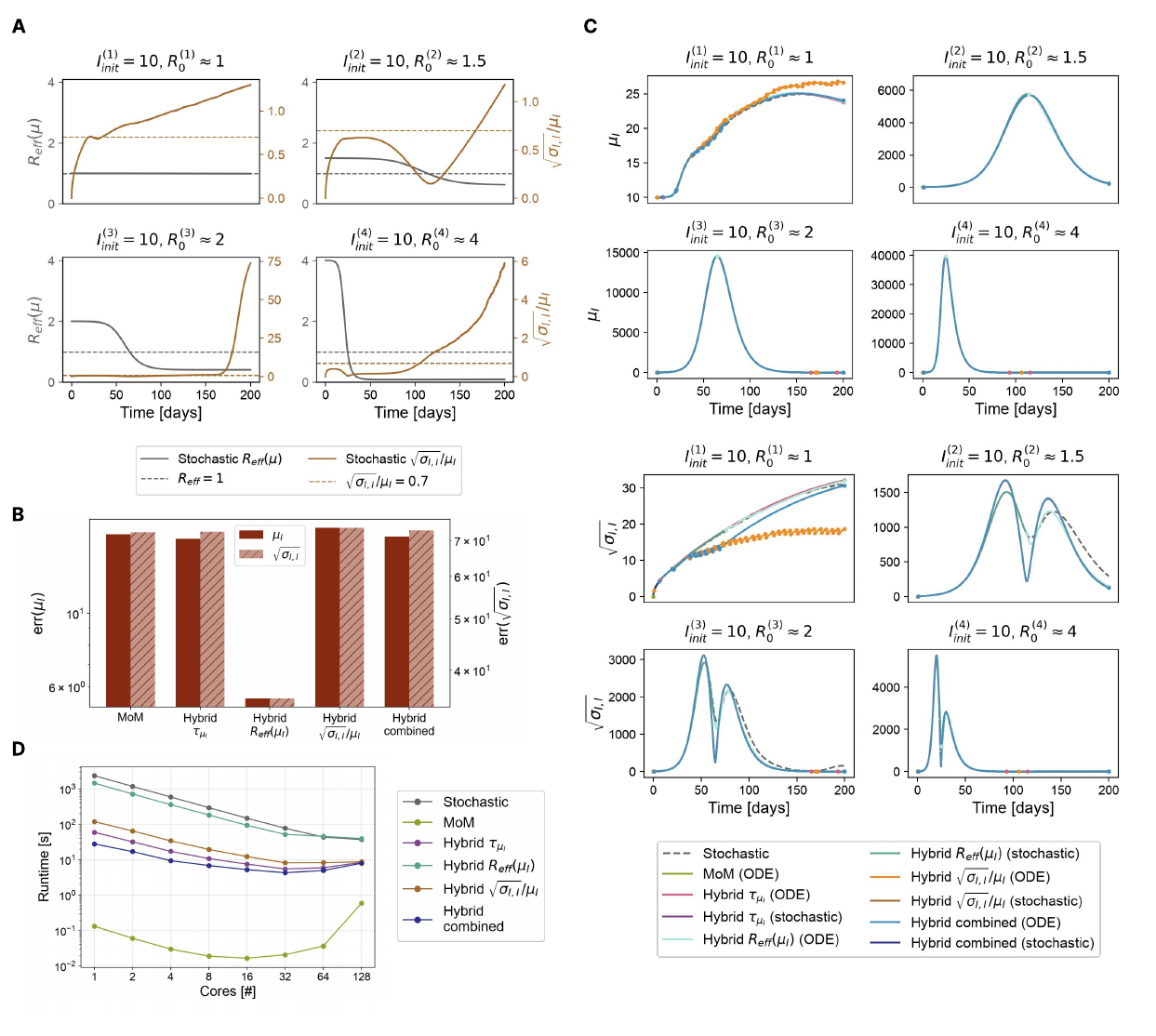}
		\caption{\textbf{Results of different condition functions for the SIR-type model with four regions and region-dependent transmission rates.} Shown are the results of $10,000$ simulations for each model. \textbf{A:} Temporal evolution of the relative standard deviation of the $I$ compartment and the reproduction number of the mean for each region. The horizontal red lines indicate the threshold values used for the corresponding conditions. \textbf{B:} Mean absolute error in mean and standard deviation of the deterministic model and the hybrid model with four different conditions. \textbf{C:} Evolution of mean (top) and standard deviation (bottom) in every region for the stochastic model, the deterministic model and the hybrid model with four different conditions. For the hybrid model the stochastic parts are displayed in a dark and the deterministic parts in a light color shade. \textbf{D:} Strong scaling of the stochastic model (Stochastic), the moment-closure approximation (MoM), and the hybrid models using the absolute threshold condition ($\tau_{\mu_I}$), the reproduction number condition ($R_{eff}(\mu_I)$), the standard deviation relation condition ($\frac{\sqrt{\sigma_{I,I}}}{\mu_I}$) and the combined condition.}
		\label{fig:supp_hybrid_results_SIR_R0}
	\end{figure}
	
	\begin{figure}[!ht]
		\centering
		\includegraphics[width=\linewidth]{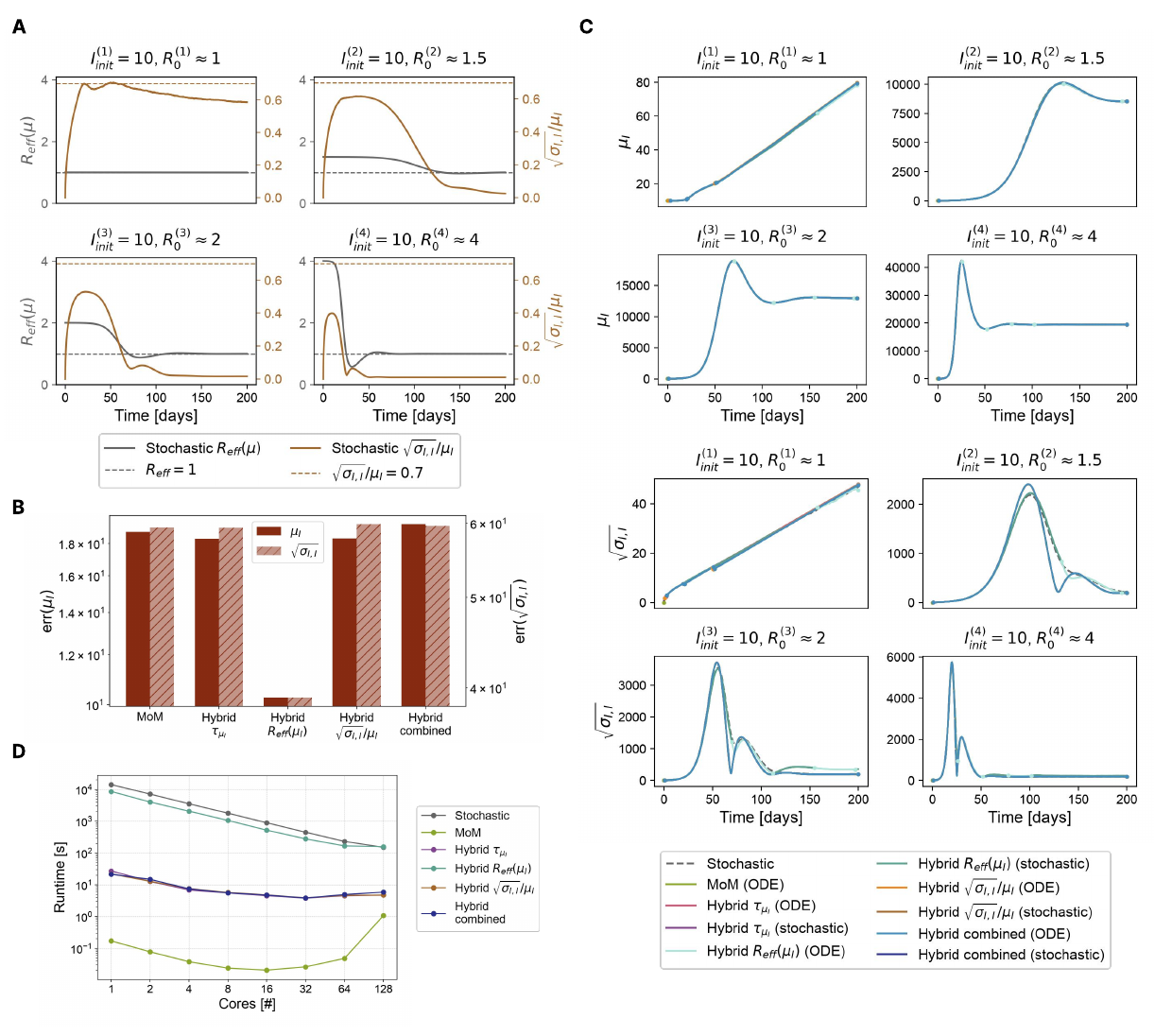}
		\caption{\textbf{Results of different condition functions for the SIRS-type model with four regions and region-dependent transmission rates.} Shown are the results of $10,000$ simulations for each model. \textbf{A:} Temporal evolution of the relative standard deviation of the $I$ compartment and the reproduction number of the mean for each region. The horizontal red lines indicate the threshold values used for the corresponding conditions. \textbf{B:} Mean absolute error in mean and standard deviation of the deterministic model and the hybrid model with four different conditions. \textbf{C:} Evolution of mean (top) and standard deviation (bottom) in every region for the stochastic model, the deterministic model and the hybrid model with four different conditions. For the hybrid model the stochastic parts are displayed in a dark and the deterministic parts in a light color shade. \textbf{D:} Strong scaling of the stochastic model (Stochastic), the moment-closure approximation (MoM), and the hybrid models using the absolute threshold condition ($\tau_{\mu_I}$), the reproduction number condition ($R_{eff}(\mu_I)$), the standard deviation relation condition ($\frac{\sqrt{\sigma_{I,I}}}{\mu_I}$) and the combined condition.}
		\label{fig:supp_hybrid_results_SIRS_R0}
	\end{figure}
	
	\begin{figure}[!ht]
		\centering
		\includegraphics[width=\linewidth]{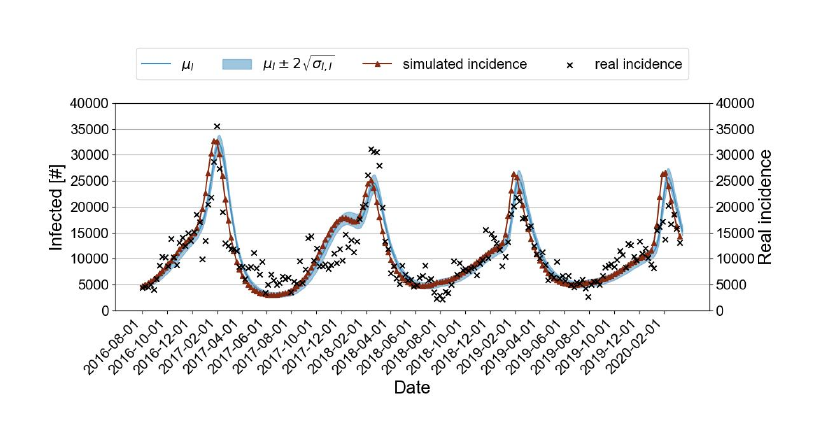}
		\caption{\textbf{Simulated influenza infected from 1st August 2016 to 9th March 2020 compared to scaled reported incidence.}}
		\label{fig:supp_influenza_fitting}
	\end{figure}
	
	\clearpage
	
	\section{Supplementary Tables}
	
	\begin{table}[!ht]
		\centering
		\footnotesize
		\renewcommand{\arraystretch}{1.4}
		\setlength{\extrarowheight}{1.5pt}
		\begin{tabularx}{1.05\textwidth}{l | Y | Y Y Y}
			\hline
			\multirow{2}{*}{Parameter} & \multirow{2}{*}{Description} & \multicolumn{3}{c}{Value per scenario} \\
			\cline{3-5}
			& & Setting 1 & Setting 2 & Setting 3 \\
			\hline\hline
			$N$ & Total population size  & Ranging from $10^3-10^7$ & $10^6$ equally distributed across regions & $10^6$ equally distributed across regions \\
			\hline
			$n_R$ & Number of regions & $1$ & Ranging from $2^0-2^4$ & Ranging from $2^0-2^4$ \\
			\hline
			$\mu_{\iota(I,k)}(t_0)$ & Mean number of initially infected per region& $10$ & $10$ & $10$\\
			\hline
			$\mu_{\iota(R,k)}(t_0)$ & Mean number of initially recovered per region& $0$ & $0$ & $0$\\
			\hline
			$\gamma$ & Recovery rate in $\mathrm{days}^{-1}$& $\frac{1}{7}$ & $\frac{1}{7}$ & $\frac{1}{7}$\\
			\hline
			$\alpha$ & Immunity waning rate in $\mathrm{days}^{-1}$& $0$ (SIR), $\frac{1}{20}$ (SIRS) & $0$ (SIR), $\frac{1}{20}$ (SIRS) & $0$ (SIR), $\frac{1}{20}$ (SIRS)\\
			\hline
			$\lambda^{(k)}$ & Transmission rate per region& $\frac{2.5\gamma}{N^{(k)}}$ & $\frac{2.5\gamma}{N^{(k)}}$ & $\frac{2.5\gamma}{N^{(k)}}$\\
			\hline
			$\kappa_z^{(k,l)}$ & Spatial transition rate & $0$ & $0$ & $10^{-4}$\\
			\hline
		\end{tabularx}
		\caption{Parameters used for scaling simulation study.}
		\label{tab:parameters_scaling_application}
	\end{table}
	
	\begin{table}[!ht]
		\centering
		\footnotesize
		\begin{tabularx}{1.05\textwidth}{l | Y | Y | l}
			\hline Parameter & Description & Value & Reference\\
			\hline\hline
			$\Delta t_{max}$ & Time interval in days at which model outcomes are saved & $0.1$ &\\
			\hline
			$n_R$ & Number of modeled regions & $1$ (global); $4$ (regional) & \cite{Buchholz_GrippeWeb_Daten_2026}\\
			\hline
			$n_{sims}$ & Number of simulations used for the stochastic model & $5{,}000$&\\
			\hline
			$\mu_{\iota(I,k)}(t_0)$ & Mean number of initially infected per region& $4{,}407$ (global); $320$ (North), $1{,}000$ (East), $1{,}400$ (South), $1{,}000$ (West) & calibrated\\
			\hline
			$\mu_{\iota(R,k)}(t_0)$ & Mean number of initially recovered per region & $198{,}228$ (global); $50{,}000$ (North), $40{,}220$ (East), $41{,}850$ (South), $37{,}700$ (West) & calibrated\\
			\hline
			$N^{(k)}$ & Initial population size per region & $835{,}771$ (global); $135{,}314$ (North), $160{,}939$ (East), $244{,}948$ (South), $294{,}569$ (West) & \cite{population_statistic}, accessed via~\cite{memilio}\\
			\hline
			$\gamma$ & Recovery rate in $\mathrm{days}^{-1}$& $1/7$ & \cite{morris_influenza_2024, ip_viral_2017}\\
			\hline
			$\alpha$ & Waning immunity rate in $\mathrm{days}^{-1}$ & $1/149$ & calibrated\\
			\hline
			$\lambda_0$ & Peak transmission rate & $2.9086e-7$ (global); $1.7992e-6$ (North), $1.5318e-6$ (East), $9.7867e-7$ (South), $8.1521e-7$ (West) & calibrated\\
			\hline
			$t^{(peak)}$ & Transmission rate peak time point per season in days& $(174, 555, 907, 1272, 1637)$ & calibrated\\
			\hline
			$\rho$ & Minimal transmission rate proportion per season & $(0.81, 0.83, 0.8, 0.78, 0.805)$ & calibrated \\
			\hline
			$\sigma$ & Seasonality standard deviation & $(20, 15, 10, 10, 12.5)$ & calibrated\\
			\hline
			$d^{(1)}=(d_{North}^{(1)}, d_{East}^{(1)}, d_{South}^{(1)}, d_{West}^{(1)})$ & First damping factor applied at 16th March 2020 & $(1,1,1,1)$ (Base), $(0.4,0.4,0.4,0.4)$ (Global NPIs, Global NPIs total lifting), $(0.4,0.4,0.3,0.4)$ (Stricter NPIs in the South), $(0.4,0.4,1,0.4)$ (No NPIs in the South) & \cite{a3m}\\
			\hline
			$d^{(2)}=(d_{North}^{(2)}, d_{East}^{(2)}, d_{South}^{(2)}, d_{West}^{(2)})$ & Second damping factor applied at 15th April 2020 & $(1,1,1,1)$ (Base), $(\frac{0.95}{0.4},\frac{0.95}{0.4},\frac{0.95}{0.4},\frac{0.95}{0.4})$ (Global NPIs, Global NPIs total lifting), $(\frac{0.95}{0.4},\frac{0.95}{0.4},\frac{0.95}{0.3},\frac{0.95}{0.4})$ (Stricter NPIs in the South), $(\frac{0.95}{0.4},\frac{0.95}{0.4},1,\frac{0.95}{0.4})$ (No NPIs in the South) & \cite{a3m}\\
			\hline
			$d^{(3)}=(d_{North}^{(3)}, d_{East}^{(3)}, d_{South}^{(3)}, d_{West}^{(3)})$ & Third damping factor applied at 2nd November 2020 & $(1,1,1,1)$ (Base), $(\frac{0.8}{0.95},\frac{0.8}{0.95},\frac{0.8}{0.95},\frac{0.8}{0.95})$ (Global NPIs, Global NPIs total lifting), $(\frac{0.8}{0.95},\frac{0.8}{0.95},\frac{0.7}{0.95},\frac{0.8}{0.95})$ (Stricter NPIs in the South), $(\frac{0.8}{0.95},\frac{0.8}{0.95},1,\frac{0.8}{0.95})$ (No NPIs in the South) & \cite{a3m}\\
			\hline
			$d^{(4)}=(d_{North}^{(4)}, d_{East}^{(4)}, d_{South}^{(4)}, d_{West}^{(4)})$ & Fourth damping factor applied at 25th November 2020 & $(1,1,1,1)$ (Base), $(\frac{0.7}{0.8},\frac{0.7}{0.8},\frac{0.7}{0.8},\frac{0.7}{0.8})$ (Global NPIs, Global NPIs total lifting), $(\frac{0.7}{0.8},\frac{0.7}{0.8},\frac{0.6}{0.7},\frac{0.7}{0.8})$ (Stricter NPIs in the South), $(\frac{0.7}{0.8},\frac{0.7}{0.8},1,\frac{0.7}{0.8})$ (No NPIs in the South) & \cite{a3m}\\
			\hline
			$d^{(5)}=(d_{North}^{(5)}, d_{East}^{(5)}, d_{South}^{(5)}, d_{West}^{(5)})$ & Fifth damping factor applied at 8th March 2021 & $(1,1,1,1)$ (Base), $(\frac{0.85}{0.7},\frac{0.85}{0.7},\frac{0.85}{0.7},\frac{0.85}{0.7})$ (Global NPIs, Global NPIs total lifting), $(\frac{0.85}{0.7},\frac{0.85}{0.7},\frac{0.85}{0.6},\frac{0.85}{0.7})$ (Stricter NPIs in the South), $(\frac{0.85}{0.7},\frac{0.85}{0.7},1,\frac{0.85}{0.7})$ (No NPIs in the South) & \cite{a3m}\\
			\hline
			$d^{(6)}=(d_{North}^{(6)}, d_{East}^{(6)}, d_{South}^{(6)}, d_{West}^{(6)})$ & Sixth damping factor applied at 3rd May 2021 & $(1,1,1,1)$ (Base), $(1,1,1,1)$ (Global NPIs), $(\frac{1}{0.85},\frac{1}{0.85},\frac{1}{0.85},\frac{1}{0.85})$ (Global NPIs total lifting), $(1,1,1,1)$ (Stricter NPIs in the South), $(1,1,1,1)$ (No NPIs in the South) & \cite{a3m}\\
			\hline
		\end{tabularx}
		\caption{Parameters used for the (regional) influenza application.}
		\label{tab:influenza_parameters}
	\end{table}
	
\end{document}